\documentclass[useAMS,usenatbib]{mnras}

\usepackage{txfonts}
\usepackage[dvips]{graphicx}
\usepackage{epsfig}
\usepackage{float}
\usepackage[usenames]{color}
\usepackage{subfig}
\usepackage{caption}

\def\lsim{\mathrel{\hbox{\rlap{\hbox{\lower4pt\hbox{$\sim$}}}\hbox{$<$}}}}
\def\gsim{\mathrel{\hbox{\rlap{\hbox{\lower4pt\hbox{$\sim$}}}\hbox{$>$}}}}

\def\apj {ApJ}
\def\aj {AJ}
\def\mnras {MNRAS}
\def\aap {A\&A}

\title[The impact of void environment on AGN]{The impact of void environment on AGN}
\author[L. Ceccarelli, F. Duplancic \& D. Garcia Lambas]{
\parbox[t]{\textwidth}{Laura Ceccarelli$^{1,3}$\thanks{E-mail: laura.ceccarelli@unc.edu.ar}, Fernanda Duplancic$^{2}$, Diego Garcia Lambas$^{1,3}$
}
\vspace*{6pt}\\
$^1$ Instituto de Astronom\'{i}a Te\'orica y Experimental (IATE), CONICET--UNC, Laprida 854, X5000BGR, C\'ordoba, Argentina. \\
$^{2}$ Departamento de Geof\'{i}sica y Astronom\'{i}a, CONICET, Facultad de Ciencias Exactas, F\'{i}sicas y Naturales, Universidad Nacional \\
de San Juan, Av. Ignacio de la Roza 590 (O), J5402DCS, Rivadavia, San Juan, Argentina\\
$^3$Observatorio Astron\'omico de C\'ordoba, Universidad Nacional de C\'ordoba, Laprida 854, X5000BGR, C\'ordoba, Argentina
}

\hypersetup{draft} 
\begin{document}
\date{\today}
\maketitle
\begin{abstract}

We study the population of active galaxies in void environment in the SDSS. 
We use optical spectroscopic information to analyze characteristics of the emission lines 
of galaxies, accomplished by WHAN and BPT diagrams. 
Also, we study WISE mid-IR colours to assess AGN activity.
We investigate these different AGN classification schemes, both optical and mid-IR, and 
their dependence on the spatial location with respect to the void centres. To this end, 
we define three regions: void, the spherical region defined by voidcentric distance relative to 
void radius (distance/r$_{\rm void}$) smaller than 0.8, comprising overdensities lesser than 
-0.9, an intermediate/transition shell region (namely void--wall) 
0.8 $<$ distance/r$_{\rm void} <$ 1.2, and a region sufficiently distant from voids, 
the field: distance/r$_{\rm void} >$ 2. 
We find statistical evidence for a larger fraction of AGN and star--forming galaxies in the 
void region, 
regardless of the classification scheme addressed (either BPT, WHAN or WISE). 
Moreover, we obtain a significantly stronger nuclear activity in voids compared to the field.
We find an unusually large fraction of the most massive black holes undergoing strong accretion 
when their host galaxies reside in voids.
Our results suggest a strong influence of the void environment on AGN mechanisms associated with  
galaxy evolution.
\end{abstract}


\section{Introduction} 
Properties of galaxies and their evolution, are known to be strongly affected by their 
local environment. For instance, stellar mass and colours of galaxies for which objects 
in high--density regions are in general more stellar massive and evolved, with a present--day low 
star-formation activity, and global redder colours \citep{dressler,blanton2005,blanton2009,kauffmann2004,cooper2006,cooper2012,balogh1998,balogh2004,gomez2003,lewis2002,koopman2004,boselli2006,jaffe2016}. 
Similarly, the large scale region where galaxies reside
also has effects on their main properties. In particular, galaxies in void surroundings 
are bluer and more actively star forming than 
similar galaxies (in both luminosity and local density) in the field 
\citep{ceccarelli_large-scale_2008}. Also, galaxies in groups at void shells have a
higher surface brightness than the corresponding galaxies in average  
global environments \citep{ceccarelli_low_2012}. 

In the extremely low dense environment typical of cosmic voids, 
the rate of merger and interactions are expected to be significantly lower than in the field or in 
groups. Besides, since the region surrounding galaxies in these environments is likely to contain large amounts of
pristine gas, these are exceptional scenarios to examine galaxy evolution.
For these reasons, the study of the galaxy population in large voids may be crucial to understand 
the processes involved in the formation and evolution of galaxies.
However, it should be considered that even though voids correspond to large volumes in the universe, covering more 
than $~40\%$ of the total volume of galaxy surveys 
\citep{hoyle_voids_2004,patiri_2006,aragoncalvo2010,pan_cosmic_2012,ceccarelli_cluesI_2013}, 
the very limited number of galaxies inside them (lower than $20\%$ the average
in density the Universe) makes it difficult to obtain samples of void galaxies suitable
for statistical analysis. This is one of the reasons why voids need to be studied statistically using 
the largest possible surveys covering wide solid angles to considerable depths.
Void statistical properties also provide invaluable information on 
higher order correlation functions \citep{white79, fry86} which can be used to probe models
for galaxy clustering \citep{croton2004} and void properties such as sizes, shapes 
and frequency of occurrence, as well as their dependence on galaxy type.
In general, void statistics may provide important clues on galaxy formation
processes and can effectively place constraints on cosmological models \citep{peebles_void_2001}.
Recent and upcoming galaxy surveys including fainter galaxies 
favours new and more detailed analyses of galaxies populating  
under--dense regions.
Several previous studies have focused on galaxies residing in cosmic voids, finding them to 
have significantly different properties than their counterparts in an average density environment. 
The luminosity function of galaxies in voids in the Sloan Digital Sky Survey
has been measured by \citet{hoyle_luminosity_2005}. These authors also studied their photometric properties
finding that in general the population of galaxies in voids is distinguished by
a fainter characteristic luminosity $L^*$. Nevertheless, the relative importance
of faint galaxies is similar to that found in the field (i.e. the faint-end
slope of the luminosity function in voids is similar to that of the field).
Spectroscopic properties of void galaxies have also been 
studied in detail \citep{rojas_spec,beygu2016}; these results indicate that 
galaxies inside voids 
have a higher star formation rates than those in denser regions, 
and are still forming stars at the present at similar rates than in the past. 
In general, void galaxies are small, blue systems, 
with late--type morphology and with a higher star formation activity 
when compared to galaxies in an average density environment 
\citep{grogin2000,parkprop2007,von_benda-beckmann_void_2008,kreckel2011,ricciardelli2014,liu2015,moorman2016,beygu2017,ricciardelli2017}. 

It has also been explored how local environment affects the active galactic nuclei (AGN) phenomena. The results 
of these studies conclude a higher active galactic nucleus occurrence in lower/medium density environments at low redshifts.  \citep{kauffmann2004,choi2009,sabater2013,Coldwell2014,Coldwell2017}. 
In addition, galaxy--galaxy interactions may produce gravitational instabilities providing gas to the most central regions of the galaxy and the growth by accretion of the central black holes. \citep[e.g,][]{Sanders1988,Storchi-Bergmann2001,Koulouridis2006,Alonso2007,Ellison2011,Satyapal2014,sabater2015,HernandezIbarra2016,Storchi-Bergmann2019}. Albeit the role of the large--scale region versus local interactions in triggering different types of AGN remain uncertain, whether quenched or enhanced, the nuclear activity is strongly related to the fuel \citep{sabater2015}. For instance optical and radio AGN are related to cold and hot gas accretion respectively \citep{argudo2016}. In this scenario, optical AGN are more likely to develop in isolated systems, with more abundant cold gas available. In this line \citet{Duplancic2021} study optical and mid--IR AGN in a sample of small galaxy systems, comprising pairs, triplets and groups with 4 an up to six galaxy members. These systems are locally isolated and given the proximity of galaxy members, galaxy--galaxy interactions and mergers are expected. The authors found a decreasing fraction of AGN galaxies with increasing number of members and highlight the important role of interactions, besides the  global/local environment dependence, in the activation of the AGN phenomenon in small galaxy systems. It is worth to notice that according to \citet{Duplancic2020} pairs and triplets in this small galaxy system sample are associated to void environments while galaxy groups are more likely to reside in void walls.

In this scheme, cosmic voids and their surrounding regions are promising candidate sites to host AGN galaxies.  
Several works report the presence of active galactic nuclei in cosmic voids. \citet{constantin2008} found that AGN are more common in voids than in walls, but only among moderately luminous and massive galaxies. 
\citet{liu2015} perform a statistical study and confirm that AGN exist in voids but with similar abundance to that in walls.  
\citet{Argudo2018} study  optical AGN in a population of isolated galaxies, and found that AGN activity appears to be “environment triggered” in quiescent isolated galaxies, where the fraction of AGN as a function of specific star formation rate and colour increases from void regions to denser large scale regions, independently of stellar mass. \citet{Amiri2019} analysed AGN in rich galaxy clusers and in voids finding that, in the local universe, the nuclear activity correlates with stellar mass and galaxy morphology and is weakly, if at all, affected by the local galaxy density.

Motivated by these previous results, in this paper we perform a statistical study of galaxies in voids and in their surroundings using SDSS data. Classification schemes of galaxies according to their emission lines provide useful insights into their star formation rate, chemical composition, and nuclear activity. In the present work we use diverse methods to classify active galaxies, tow are based on optical spectral features by using the BPT  \citep{baldwin1981} and WHAN \citep{cidfernandes2011} diagnostic diagrams, the third method uses infrared photometry which is sensitive to optically obscured AGN.
Our study is centered on analysis of galaxy properties related to nuclear activity, and takes into account the relative position with respect to the void centre.

This paper is organised as follows.  
In section 2 we introduce the catalogue and define the galaxy 
samples we will use in our statistical measurements.
In section 3 we briefly describe the void identification algorithm 
and show some properties of the resulting samples of voids.
We examine and compare the different diagnostic diagrams 
where we consider separately galaxies in void walls, interiors and in the field in section 4.  
 We analyse the fraction of galaxies according to classes at different distances to void in section 5 and 
we examine the nuclear activity of galaxies 
prevailing at voids in section 6.
In section 7 we present a summary and finally, in section 8, we give a brief discussion with  possible interpretations
of our results.

\section{Data}
This work is based on the Sloan Digital Sky Survey data release 7 \citep[\textsc{SDSS-DR7,}][]{abazajian2009}. 
Due the small number of galaxies in underdense region, in order to increase the data we have included in our analysis diverse AGN selection criteria. 

\subsection{Galaxy catalogue: SDSS}\label{sec:datavls}

Data used to develop this work have been obtained from catalogues through $\rm SQL$ queries in the  publicly  available  Catalog  Archive  Server (CAS)\footnote{http://skyserver.sdss.org/casjobs/}. Photometric properties were taken from \texttt{Galaxy} view and spectroscopic information from \texttt{SpecObjAll} table. 

The data used for void identification is extracted from the Main
Galaxy Sample \citep{strauss_spectroscopic_2002} of the
\textsc{SDSS-DR7}.
The SDSS contains CCD imaging data in five photometric bands
\citep[$ugriz$,][]{fukugita_sloan_1996, smith_ugriz_2002}.
The \textsc{SDSS-DR7} spectroscopic catalogue comprises in this
release 929,555 galaxies with a limiting petrosian aparent magnitude of
\mbox{$r_{mag}~\leq~17.77$}.\\
Considering that the catalogue used here are apparent magnitude limited surveys, 
the brightest galaxies are present in the whole catalogue while 
the faintest galaxies are only recorded at small distances.
As a consequence, a luminosity bias appears which can be overcome by constructing volume 
limited samples. 

For a given maximum redshift $z_{lim}$ SDSS is complete for galaxies 
brighter than 
   \begin{equation}
   {\rm Mr_{lim}}=r_{lim}-25-5*{\rm log}(d_{lim}),
   \label{eq:blim}
   \end{equation}
where $r_{lim}$ is the flux completeness limit and $d_{lim}$ is the luminosity distance obtained by $d_{lim}= (c/H_0) \int_{0}^{z} \! (\Omega_m(1+z)^3+\Omega_{\Lambda})^{1/2} \, dz $, where 
c is the speed of light in vacuum and $H_0$ is the Hubble constant at present which
can be expressed in terms of a dimensionless Hubble parameter $h$, $H_0=100\  h\ \rm km s^{-1}/ \rm Mpc $, 
$\Omega_m$ is the density parameter and $\Omega_{\Lambda}$ the cosmological constant, for which we adopt the values 0.26 and 0.74 respectively. 

We construct a SDSS volume limited sample with galaxies brighter than 
$R_{lim}$ corresponding to redshifts $z<z_{lim}=$0.15. 
We calculated galaxy absolute magnitudes  using the relation ${\rm M_r}=r_{mag} -25-5*{\rm log}(d_l)-k_r$, 
where $r_{mag}$ is the r band apparent magnitude, $d_l$ is the luminosity distance calculated according equation \ref{eq:blim} and $k_{r}$ the k-correction. 
In order to correct the absolute magnitudes we use the SDSS k-corrections developed by \citet{omill2011MNRAS}
and we restrict our analysis to galaxies with $r$-band magnitudes in the range $14<r_{mag}<17.77$. This range avoids saturated stars as well as assures spectroscopic completeness in SDSS Main Galaxy Sample \citep{strauss_spectroscopic_2002}. We apply the apparent magnitude cut after Galactic extinction correction, in order to obtain an uniform extinction-corrected sample. 

This sample is used to identify voids in the galaxy distribution, as well as to analyse galaxies associated to these structures.


\subsection{AGN data} \label{sec:agnsample}

In order to identify AGN host galaxies in our sample, we use three different methodologies. Two methods are based on spectroscopic optical data using emission-line ratios, and the third one uses infrared photometry making it more sensitive to optically obscured AGN.

 For the optical  AGN selection, we use the publicly available SDSS emission-line fluxes taken from the MPA/JHU VAGC\footnote{Available in CAS \texttt{galSpecIndx, galSpecInfo, galSpecLine and galSpecExtra} tables}. The method for emission-line measurement is detailed in \citet{Tremonti2004} and \citet{Brinchmann2004}. 
To accomplish the infrared AGN selection, we used the methodology described in \citet{Assef2018} based on the Wide-field Infrared Survey
Explorer \citep[WISE\footnote{The cross-matched WISE data is available in CAS \texttt{WISE\_allsky} and  \texttt{WISE\_xmatch} tables}; ][]{Wright2010}. This survey comprises mid-IR images of the entire sky in four bands, centered at 3.4, 4.6, 12, and 22$\mu m$, hereafter w1, w2, w3, and w4, respectively. The derived mid-IR AGN catalogues are based on the AllWISE Data Release \citep{Cutri2013}.

\section{Voids in the galaxy distribution}

In order to identify voids in the galaxy distribution we adopt the void finding algorithm described in \citet{padilla_spatial_2005} and 
\citet{ceccarelli_voids_2006},
which produces a large number of candidate void centres at random throughout the
galaxy catalogue and then tests the density within spheres centred on these random
points for a wide range of sphere radii.
For each centre, the largest sphere that satisfies a low density criterion 
is selected and becomes a void candidate (note that most of the candidate centres will not
satisfy the low density condition).
In a final step, the overlaps between spheres are eliminated by removing the smaller spheres superimposed on the larger ones. We keep the largest sphere in this process  corresponding to a void in our identification scheme.

We apply the void finding algorithm previously described to the observational sample described in section \ref{sec:datavls} 
and identify 323 voids within a radius range 8$<$r$_{\rm void}$/h$^{-1}$Mpc$<$30 (with mean 15.6$\pm$0.2 h$^{-1}$Mpc) enclosing an integrated mass underdensity $ \delta \le -0.9 $.
For reference, several properties of voids identified with similar methods in numerical simulations, which include their radial expansion velocities,
distributions of void radius, etc., are discussed in detail in \citet{ceccarelli_cluesI_2013,paz_clues_2013}. 
We notice that low redshift samples with a low luminosity cutoff, comprise large numbers of small voids and very few large ones. On the contrary, galaxy samples reaching higher redshifts mainly contain large voids.  These differences are a combination of the sample volume and
the limiting magnitude ($R_{lim}$) which has associated shot noise.

   \begin{figure*}
   \epsfxsize=0.9\textwidth
      \centerline{\epsffile{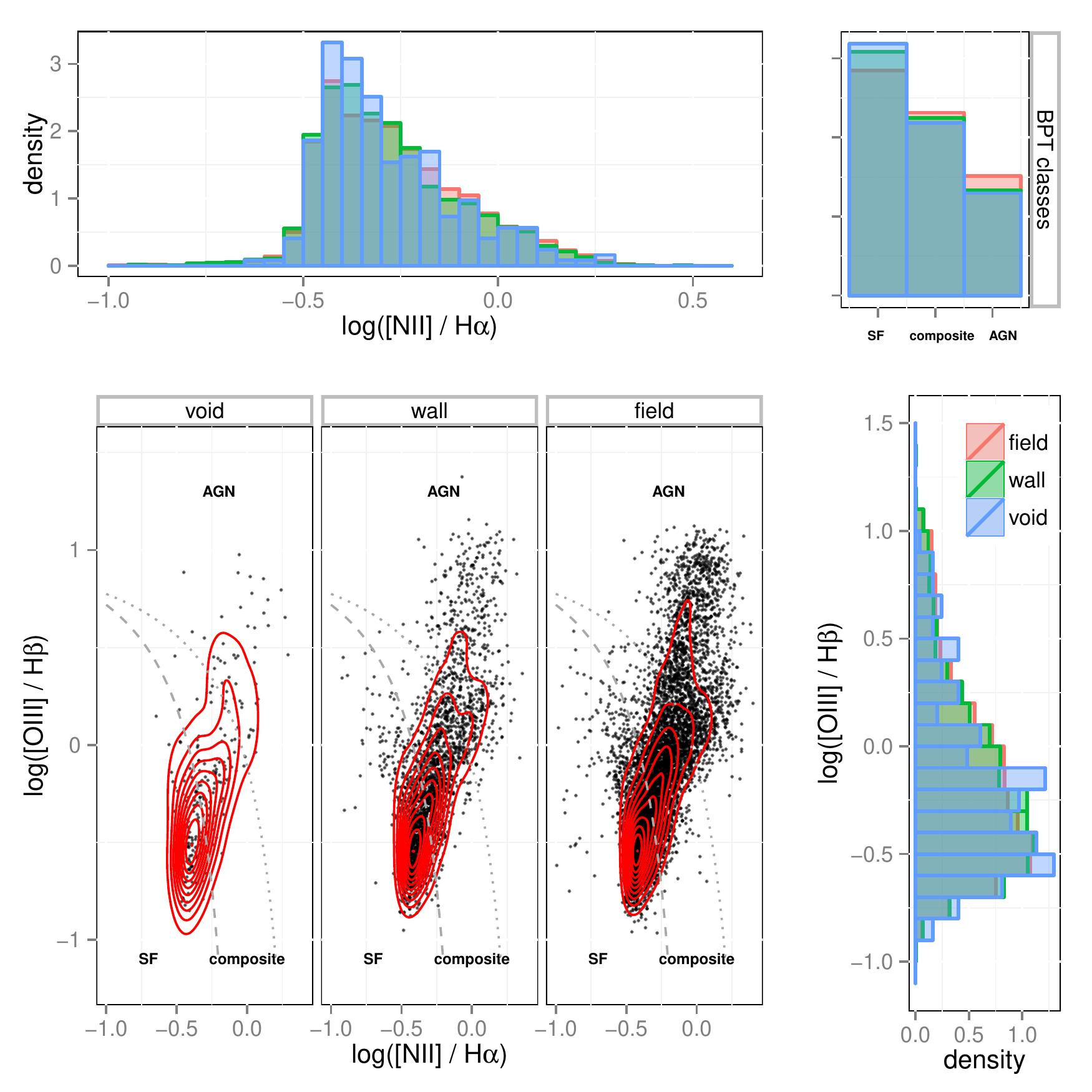}
   }
   \caption{
   BPT diagram and corresponding marginal distributions of [NII]/H$\alpha$ and [OIII]/H$\beta$ for galaxies in voids (left scatter plot and blue shadow), walls (middle scatter plot and green shadow) and field (right scatter plot and red shadow). Red lines indicate isodensity levels and the gray lines demarcate the regions of the diagram corresponding to the different classes of galaxies which are indicated in the figure. 
   The small upper right panel shows the relative population of the different classes of galaxies in voids (blue), walls (green) and field (red).
      }
   \label{fig:bptagn}
   \end{figure*}
   
   \begin{figure*}
   \epsfxsize=0.9\textwidth
      \centerline{\epsffile{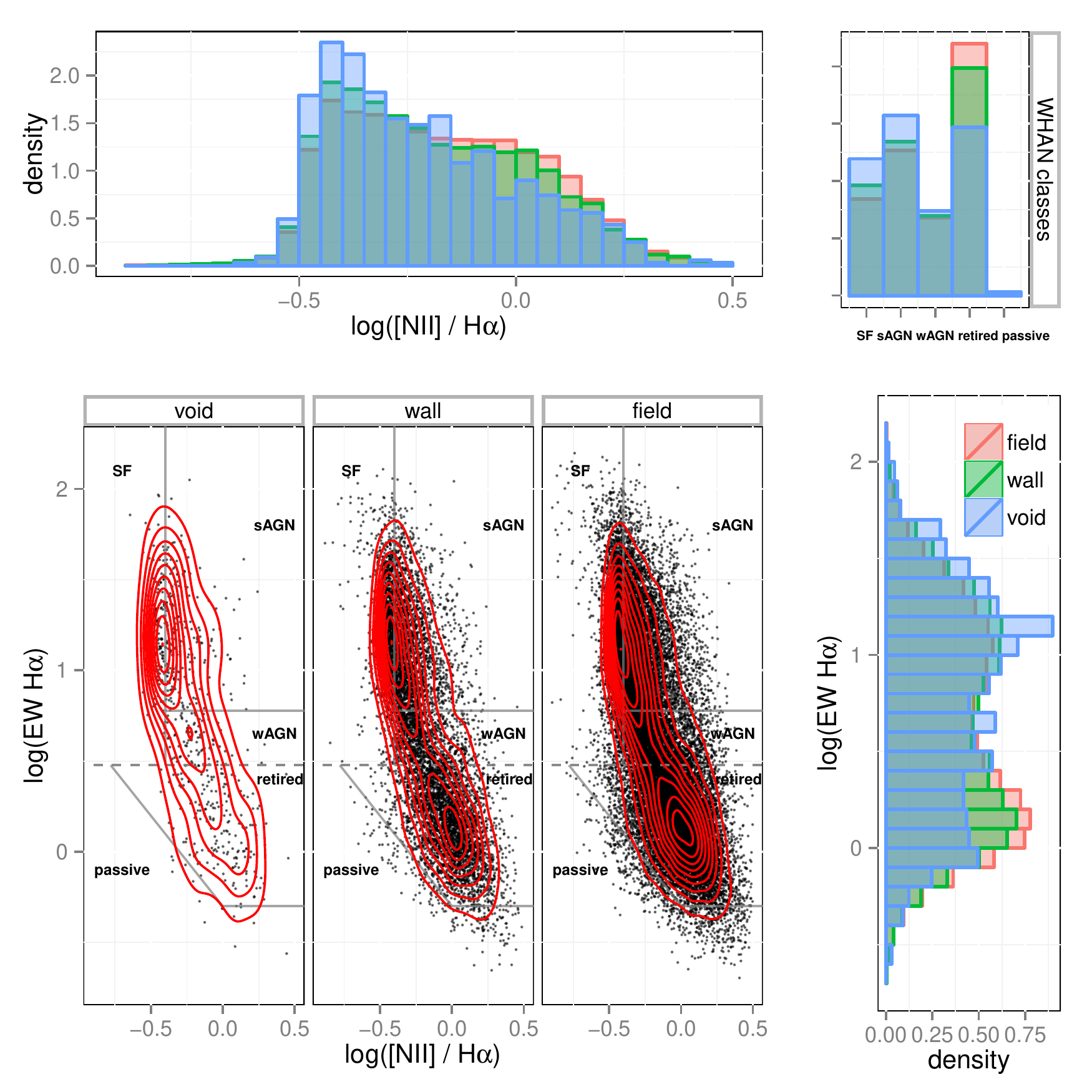}
   }
   \caption{
   WHAN diagram and corresponding marginal distributions of [NII]/H$\alpha$ and EW(H$\alpha$) for galaxies in voids (left scatter plot and blue shadow), walls (middle scatter plot and green shadow) and field (right scatter plot and red shadow). Red lines indicate isodensity levels and the gray lines demarcate the regions of the diagram corresponding to the different classes of galaxies which are indicated in the figure. 
   The small upper right panel shows the relative population of the different classes of galaxies in voids (blue), walls (green) and field (red).
       }
   \label{fig:whangals}
   \end{figure*}
   \begin{figure*}
   \epsfxsize=0.9\textwidth
   \centerline{\epsffile{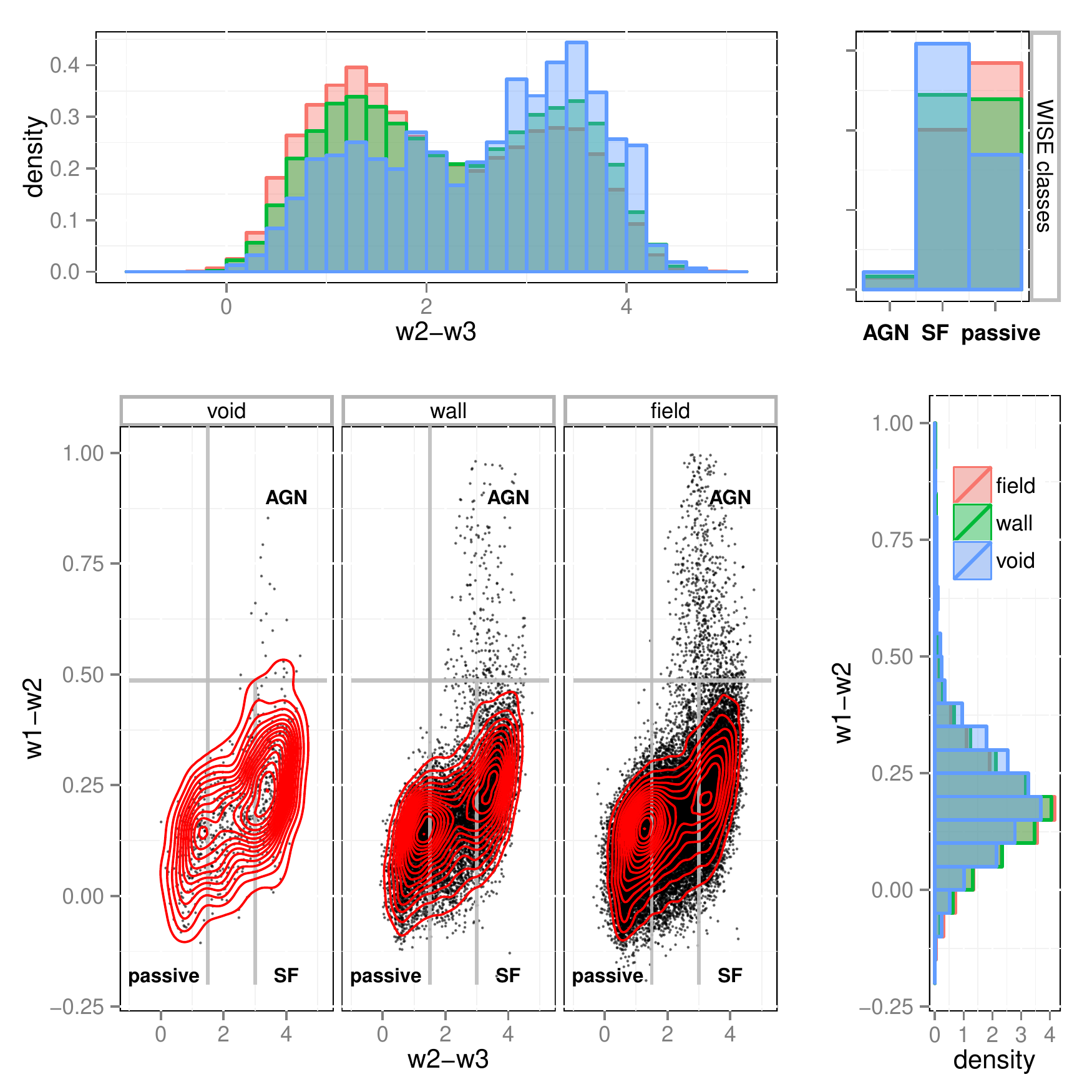}
   }
   \caption{WISE colour--colour diagram and corresponding marginal distributions of (w2-w3) and (w1-w2) colours for galaxies in voids (left scatter plot and blue shadow), walls (middle scatter plot and green shadow) and field (right scatter plot and red shadow). Red lines indicate different isodensity levels and the gray lines demarcate the regions of the diagram corresponding to the different classes of galaxies which are indicated in the figure. 
   The small upper right panel shows the relative population of the different classes of galaxies in voids (blue), walls (green) and field (red).
       }
   \label{fig:wisegals}
   \end{figure*}

%

The aim of this work is to investigate whether large--scale underdense regions are particular locations for AGN compared to other environments. 
To analyze this possibility, we start by assigning each galaxy a large-scale environment: either void interior (hereafter void), void--wall (hereafter wall) or field.
We refer as void galaxies to those objects residing within void--centric distances smaller than 0.8 void radius enclosing an integrated mass underdensity $ \delta \le -0.9 $, void--wall or wall galaxies comprise those at voidcentric distances between 0.8 and 1.2 r$_{\rm void}$, with typical underdensities $\delta \thickapprox -0.8 $ and with large expansion velocity with respect to the void centre. Finally we consider a sample of field galaxies, corresponding to more typical environment, as those located far from the closest void (distance/r$_{\rm void} >$ 2). We notice that galaxies at distances between 1.2 and 2 in r$_{\rm void}$ units correspond to an intermediate--field close to the void--wall regions, for this reason, these galaxies are not used for comparisons, and use rather, more distant galaxies from the void centers.  

Throughout this work we have analyzed samples of SDSS galaxies selected according to their total r--band luminosity and redshift. We have adopted -23$<\rm M_r<$-20.5 and 0.05$<$z$ <$ 0.15, obtaining with these restrictions a total number of 188204 galaxies. 
The limiting magnitude was chosen based on the characteristic galaxy luminosity of the SDSS luminosity function \citep{blanton2003lf}. The lower limit in the redshift range prevents small fixed-size apertures from affecting galaxy properties derived from the fibre spectra \citep[][]{Coldwell2017}.
Considering the large-scale environment categories defined previously our subsamples comprise 782 galaxies in voids, 11558 in walls, and for comparison, there are 96770 in the field region distant from void centres. 
Table \ref{table:samples} summarizes the main characteristics of the galaxy subsamples selected according to their large--scale environment (first to fourth columns).
\\
We recall that both redshift and r-band absolute magnitude distributions of galaxies in the different large--scale regions considered are remarkably similar (p$>$0.05 for a Kolmogorov-Smirnov test). Thus, our results are not biased neither to redshift nor luminosities differences across the samples residing in voids, walls and field.
This is also verified with the galaxy subsamples used in the different studies carried out throughout this work.

\section{AGN in Voids}
We have used different spectroscopic and photometric diagnostic diagrams applied to the galaxies of the subsamples classified according to our large-scale environment definition. We have considered in this work three cases: BPT and WHAN emission line diagnostic diagrams as well as WISE colour-colour diagram for the sub-samples of galaxies in voids, walls and field. This analysis is aimed to analyze possible differences in the diagram output according to the host galaxy location with respect to the voids.  We recall that for the following study galaxies were selected with the redshift and luminosity restrictions 0.05$<z<$0.15 and -23.0$< \rm M_r <$ -20.5. 

\subsection{BPT emission line diagnostic diagram}\label{sec:bpt}
The most widely used method to identify optical AGN relies on diagnostic diagrams originally proposed by \citet{baldwin1981}, hereafter BPT. These diagrams allow the separation of type 2 AGN from normal, star--forming galaxies, using emission-line ratios. 
In the present work, we use the emission line classification method provided by the MPA-JHU spectroscopic analysis based on \citet{Brinchmann2004}. These authors consider galaxies with  [OIII]$\lambda$5007 (hereafter [OIII]),  [NII]$\lambda$6583 (hereafter [NII]), H$\alpha$ and H$\beta$ emission lines with a signal to noise ratio S/N$>$3 and classify galaxies according to their relative position in the [OIII]/H$\beta$ versus [NII]/H$\alpha$ BPT diagram. Star--forming galaxies (SF) correspond to objects placed below the \citet{kauffmann2003} threshold. Composite galaxies reside between \citet{kauffmann2003} and \citet{Kewley2001} boundaries. These Composite spectra are expected to exhibit a mix of star forming and AGN emission features \citep{Kewley2006}. They consider the AGN population consisting of all galaxies above the \citet{Kewley2006} boundary threshold. 
In the present work we exclude low  ionization  nuclear  emission  line  regions  (LINERs),  since the AGN category of LINERs  is  still on  debate \citep[e.g.][]{Maoz2005,Yan2012,Singh2013,Herpich2016,Coldwell2017}.

BPT diagnostic diagrams for SDSS galaxies according to their large--scale environment are shown in the three left lower panels of figure \ref{fig:bptagn}. 
Each panel corresponds to void, wall and field as indicated in the figure. 
It can be clearly seen the large  difference of the number of galaxies in each region, consistent with their definition (a density of the order of 10--20\% of the field is expected in the walls). 
Nevertheless, this contrasts with the distribution of galaxies in the diagnostic diagram which appears with similar numbers in all large--scale regions. \\
Fig. \ref{fig:bptagn} also shows the corresponding marginal distributions in the upper left and lower right panels, where each shading colour corresponds to the large-scale region (blue, green, and red for voids, walls, and field, respectively). Both marginal distributions look similar for the three large--scale regions that we are comparing. It can be noted, however, a slight tendency for void galaxies to have smaller log([NII]/H$\alpha$) and a more concentrated distribution (left upper panel). \\
In addition, the small upper right panel shows the relative proportions of the 3 classes of galaxies represented in the BTP diagram (star forming, composite and AGN, indicated on the x-axis) in voids, walls and field.

\subsection{WHAN emission line diagnostic diagram}\label{sec:whan}

Although the BPT diagram  is the most widely used method for optical AGN classification, it is no suitable to effectively disentangle the so-called retired galaxy population from weak AGN types. Retired galaxies are those objects without significant [OIII] emission and with ionization mechanisms usually explained by Hot Low-Mass Evolved Stars \citep[HOLMES;][]{Binette1994}. 
\citet{cidfernandes2011} considers that the WHAN diagram, i.e. the equivalent width  of H$\alpha$ (EW(H$\alpha$)) vs [NII]/H$\alpha$ can be successfully used to separate the retired galaxy population from weak AGN types. Moreover, since only two lines are involved in this diagram, signal to noise restrictions are lower and thus a higher percentage of emission line galaxies can be classified compared to the BPT diagnostic diagram (see Table \ref{table:samples}). For the WHAN diagram we consider a signal to noise  S/N$>$3 for the two lines used in this approach and
following \citet{cidfernandes2011}, we identify strong AGN as galaxies with log([NII]/H$\alpha$)$>$-0.4 and EW(H$\alpha$)$>$6 and weak AGN fulfilling  log([NII]/H$\alpha$)$<$-0.4 and 3$<$EW(H$\alpha$)$<$6. Also, we select star--forming (SF) galaxies (log([NII]/H$\alpha$)$<$-0.4 and EW(H$\alpha$)$>$3) and retired/passive objects (EW(H$\alpha$)$<$3). 

In Fig. \ref{fig:whangals} we present the WHAN diagnostic diagrams where each of the subpanels contained in the lower left panel corresponds to galaxies in voids, walls and field, as indicated, and analogously to Fig. \ref{fig:bptagn}.
The black straight lines delimit the different WHAN galaxy classes as indicated in the figure, the red lines correspond to isodensity curves.
Besides the difference in the number of galaxies in each region, the diagrams are similar. However, a detailed inspection of the contour lines show that they differ in voids, specifically in the region of high EW(H$\alpha$), which corresponds to star forming galaxies and strong AGN.
This can be seen more clearly in the marginal distributions shown in the upper left and lower right panels of figure \ref{fig:whangals}, note that the filled colours of the histograms correspond to voids, walls or field as indicated in the lower right panel.
It can be seen that both distributions are remarkably different for void galaxies compared to wall and field galaxies. 
Regarding to log(NII/H$\alpha$) (left upper panel), the value at which the maximum of the distribution occurs is similar for voids, wall and field. However, this quantity tend to be smaller in voids, presenting a more concentrated distribution towards smaller values (more negative skew value in voids than in walls and field).  
As for the distribution of EW(H$\alpha$) (right bottom panel), this plot also shows notable differences between galaxies in voids (blue shading) with respect to those in walls and field (green and red respectively). As it can be seen in the histograms, the EW(H$\alpha$) distributions show a remarkable bimodal behavior for galaxies in the walls and field, while this behavior is negligible for galaxies in voids. 

Note the difference between the upper left panels of Figures \ref{fig:bptagn} and \ref{fig:whangals}, which represent the marginal distribution of the same quantity. This difference is due to the fact that the galaxy samples are not the same, since the number of galaxies eligible to qualify with the WHAN diagnostic diagram is greater than with the BPT. Consequently, the subsamples of galaxies in Fig. \ref{fig:bptagn} are contained in those in Fig. \ref{fig:whangals} and the latter are more numerous.

In the small upper right panel of Fig. \ref{fig:whangals} we show  
the relative population of the different galaxy classes obtained from the WHAN diagram in voids (blue), walls (green) and field (red), the different classes are indicated on the horizontal axis. 
As it can be seen, SF and AGN galaxies are more likely found in large-scale low-density environments, voids and walls, than in the field. On the contrary, the opposite occurs with retired galaxies which are more frequently found in the field.

\subsection{WISE colour-colour diagnostic diagram}

Observations show that a considerable fraction of the optical AGN emission can be absorbed by the dust surrounding the central super massive black hole. This radiation is re-emitted at IR wavelengths and therefore, mid-infrared surveys such as WISE are mostly relevant to identify optically obscured AGN. There are several WISE colour diagnostics criteria to select AGN \citep{Mateos2012,Stern2012,Assef2013}. Most of these criteria are based on the fact that AGN present a (w1-w2) colour redder than non-active galaxies since warm dust emission is more important than the light of the old stellar population of the host galaxy \citep{Stern2012,Assef2013}. Moreover, at low redshifts, the (w1-w2) colour index is less affected by extinction so that these criteria based on (w1-w2) are efficient in selecting obscured AGN.

\ In this work we have adopted the  75\%  reliability criteria presented in \citet{Assef2018} for the AGN selection based on WISE data. Therefore, active objects must fulfil

\begin{equation}
 (\rm w1-w2) > \left\lbrace
 \begin{array}{ll}
  0.486 \textup{ exp}[0.092(\rm w2-13.07)^2] & \textup{if } \rm w2>13.07 \\
  0.486 &  \textup{if } \rm w2\le13.07
 \end{array}
 \right.
\end{equation}

 Under this condition, it is expected that more than 75\% of the objects selected have bolometric luminosities dominated by an AGN \citep[for a detailed description see][]{Assef2018}.

Mid-IR WISE (w2-w3) colour can also be used to identify star--forming and passive galaxies \citep[][]{Cluver2014, Herpich2016,Jarrett2019}. In this work we adopt the methodology of \citet{Jarrett2017} and select star--forming galaxies (SF) as non-AGN with (w2-w3)$>$3. In this line we also consider as passive objects galaxies with (w2-w3)$<$1.5.

Finally, we have explored galaxy colour--colour diagrams derived from WISE magnitudes for galaxies in our different large scale environments. The results are shown in Fig. \ref{fig:wisegals}.
The three left bottom panels displayed the diagrams for 
galaxies subsamples in voids, walls and field as it is indicated in the top of each panel. 
Black points correspond to galaxies (we show the full subsamples) and red lines indicates 
the iso-density contours. The left upper panel and the right bottom panel contains the marginal colour index distributions where different colours correspond to different large scale environment 
(blue, green and red correspond to void, wall and field respectively).
The bimodality of the (w2-w3) colour distribution appears on all large scale regions. 
However some distinctions can be noticed from a more detailed inspection to the figure. 
The (w2-w3) distributions in voids and field differ remarkably and the relative importance of local maximums inverts from voids to field 
(blue and red histograms in the left upper panel). On the other hand, (w1-w2) distribution are more similar with a slight tendency for galaxies in voids to have higher (w1-w2) values than walls and field counterparts.\\
Similarly to figures \ref{fig:bptagn} and \ref{fig:whangals}, the small upper right panel shows the relative population of the different classes of galaxies obtained with the corresponding diagram in voids, walls and field.
Although AGN classified from the WISE diagram are very few, it can be seen a similar behavior  to that obtained with the WHAN diagram, where passive galaxies dominate in the field, and both SF and AGN populate regions of low global density. 

\begin{table*}
  \begin{center}
   \caption{Summary of the main characteristics of galaxy samples and the resulting number of galaxies in each sample. Galaxies are selected in the (0.05, 0.15) and (-23.0, -20.5) redshift and absolute magnitude ranges respectively}
    \begin{tabular}{cccccccccccccc}
      \hline
      \hline
           \noalign{\vglue 0.2em}
       \multicolumn{14}{ |c| }{sample selection criteria} \\ 
            \noalign{\vglue 0.2em}
      \hline
      \noalign{\vglue 0.2em}
     large-scale & distance & mean & full & \multicolumn{3}{|c|}{BPT diagram} & \multicolumn{4}{c|}{WHAN diagram}& \multicolumn{3}{c}{WISE diagram}\\
      \noalign{\vglue 0.2em}
            \noalign{\vglue 0.2em}
   region & [R$_{void}$] & r$_{mag}$ & sample & SF & AGN & composite & SF & sAGN & wAGN & ret/pas & SF & AGN & passive  \\
      \noalign{\vglue 0.2em}
      \hline
      \noalign{\vglue 0.2em}
      void & d$<$0.8 & -20.89$\pm$0.02 & 782 & 118 & 48 & 81 & 149 & 192 & 89 & 181 & 297 & 21 & 156 \\ 
      \noalign{\vglue 0.2em}
      \cline{1-14}
      \noalign{\vglue 0.2em}
      wall & 0.8$<$d$<$1.12 & -20.93$\pm$0.01 & 11558 & 1283 & 555 & 936 & 1648 & 2296 & 1175 & 3338 & 3480 & 231 & 3354 \\ 
      \noalign{\vglue 0.2em}
      \cline{1-14}
      \noalign{\vglue 0.2em}
      field & d$>2$ & -20.940$\pm$0.003 & 96770 & 9152 & 7481 & 5304 & 12006 & 17522 & 9586 & 30232 & 25264 & 2003 & 33644\\ 
      \noalign{\vglue 0.2em}
       \hline
      \hline
    \end{tabular}
    \label{table:samples}
  \end{center}
\end{table*}

As a summary of this section, we present in Table \ref{table:samples} the number of galaxies in the different categories according to the BPT, WHAN and WISE diagnostic diagram classifications.

\subsection{Comparison of diagnostic diagrams}

\begin{table*}
  \begin{center}
   \caption{Comparison between AGN diagnostic diagrams for galaxies in voids, walls and field. Percentages of cross--classes in the different diagnostic diagram used to classified galaxies.}
    \begin{tabular}{cccccccc}
      \hline
      \hline
           \noalign{\vglue 0.2em}
       \multicolumn{8}{ |c| }{BPT comparison} \\ 
            \noalign{\vglue 0.2em}
      \hline
      \noalign{\vglue 0.2em} 
     BPT class & large scale region &  WHAN SF & WHAN AGN & WHAN retired/passive & WISE AGN & WISE SF & WISE passive \\
      SF & void  & 88 $\pm$ 4 & 12 $\pm$ 1 & --- & 1.2 $\pm$ 0.3 & 93 $\pm$ 4 & 0.3 $\pm$ 0.1 \\ 
      SF & wall  & 85 $\pm$ 2 & 15.1 $\pm$ 0.5 & 0.02 $\pm$ 0.02 & 1.2 $\pm$ 0.1 & 92 $\pm$ 2 & 0.27 $\pm$ 0.05 \\ 
      SF & field & 86 $\pm$ 1 & 14.2 $\pm$ 0.3 & 0.08 $\pm$ 0.02 & 1.30 $\pm$ 0.07 & 92 $\pm$ 1 & 0.40 $\pm$ 0.04\\
      Composite & void  &  4 $\pm$ 1 & 90 $\pm$ 10 & 5 $\pm$ 1 & 5 $\pm$ 1 & 73 $\pm$ 8 & 1.7 $\pm$ 0.8 \\ 
      Composite & wall  & 3.3 $\pm$ 0.4  & 90 $\pm$ 4 & 6.5 $\pm$ 0.6 & 5.5 $\pm$ 0.6 & 71 $\pm$ 3 & 2.8 $\pm$ 0.4  \\ 
      Composite & field & 3.3 $\pm$ 0.2 & 90 $\pm$ 2 & 6.9 $\pm$ 0.4 & 5.0 $\pm$ 0.3 & 70 $\pm$ 2 & 3.8 $\pm$ 0.3 \\
      AGN  & void  & 2 $\pm$ 1 & 86 $\pm$ 15 & 11 $\pm$ 4 & 28 $\pm$ 7 & 38 $\pm$ 8
 & 7 $\pm$ 3 \\ 
      AGN  & wall  & 1.2 $\pm$ 0.3 & 76 $\pm$ 5 & 23 $\pm$ 2 & 20 $\pm$ 2 & 37 $\pm$ 3 & 12 $\pm$ 1 \\ 
      AGN  & field & 1.1 $\pm$ 0.2 & 74 $\pm$ 3 & 24 $\pm$ 1 & 19 $\pm$ 1 & 36 $\pm$ 2 & 15 $\pm$ 1\\
      \noalign{\vglue 0.2em}
            \noalign{\vglue 0.2em}
          \hline
      \hline
               \noalign{\vglue 0.2em}
       \multicolumn{8}{ |c| }{WHAN comparison} \\ 
            \noalign{\vglue 0.2em}
      \hline
      \noalign{\vglue 0.2em} 
     WHAN class & large scale region &  BPT SF & BPT composite & BPT AGN & WISE AGN & WISE SF & WISE passive \\ 
      SF  & void   & 77 $\pm$ 4  & 0.7 $\pm$ 0.2   & 0.15 $\pm$ 0.09 & 1.3 $\pm$ 0.3 & 93 $\pm$ 4 & 0.3 $\pm$ 0.1\\ 
      SF  & wall   & 73 $\pm$ 1  & 0.64 $\pm$ 0.07 & 0.11 $\pm$ 0.03 & 1.1 $\pm$ 0.1 & 92 $\pm$ 2 & 0.30 $\pm$ 0.05\\ 
      SF  & field  & 72.5 $\pm$ 0.8 & 0.63 $\pm$ 0.04 & 0.11 $\pm$ 0.02 & 1.16 $\pm$ 0.06 & 91 $\pm$ 1 & 0.42 $\pm$ 0.03\\
      AGN & void   & 19 $\pm$ 2 & 29 $\pm$ 2 & 11 $\pm$ 1 & 4.9 $\pm$ 0.8 & 76 $\pm$ 5 &  0.7 $\pm$ 0.3\\ 
      AGN & wall   & 19.4 $\pm$ 0.7 & 26.4 $\pm$ 0.8 & 10.3 $\pm$ 0.4 & 4.6 $\pm$ 0.3 & 76 $\pm$ 2 & 1.2 $\pm$ 0.1 \\ 
      AGN & field  & 18.0 $\pm$ 0.4 & 25.7 $\pm$ 0.5 & 10.9 $\pm$ 0.3 & 4.6 $\pm$ 0.2 & 74 $\pm$ 1 & 1.63 $\pm$ 0.09\\
      retired/passive  & void & ---          & 3.3 $\pm$ 0.9 & 2.7 $\pm$ 0.8 & 0.2 $\pm$ 0.2 & 9 $\pm$ 2 & 35 $\pm$ 4 \\ 
      retired/passive  & wall & 0.04 $\pm$ 0.03 & 2.6 $\pm$ 0.2 & 4.1 $\pm$ 0.3 & 0.26 $\pm$ 0.07 & 9.7 $\pm$ 0.5 & 36 $\pm$ 1 \\ 
      retired/passive  & field & 0.12 $\pm$ 0.02 & 2.3 $\pm$ 0.1 & 4.2 $\pm$ 0.2 & 0.13 $\pm$ 0.02 & 9.0 $\pm$ 0.3 & 37.4 $\pm$ 0.7\\
      \noalign{\vglue 0.2em}
            \noalign{\vglue 0.2em}
          \hline
      \hline
                     \noalign{\vglue 0.2em}
       \multicolumn{8}{ |c| }{WISE comparison} \\ 
            \noalign{\vglue 0.2em}
      \hline
      \noalign{\vglue 0.2em} 
     WISE class & large scale region &  BPT SF & BPT composite & BPT AGN & WHAN SF & WHAN AGN & WHAN retired/passive \\ 
      SF & void  &    59 $\pm$ 3  & 9 $\pm$ 1  & 1.8 $\pm$ 0.3 & 67 $\pm$ 3 & 29 $\pm$ 2 & 1.7 $\pm$ 0.3 \\ 
      SF & wall  &   54 $\pm$ 1  & 9.3 $\pm$ 0.3  & 2.2 $\pm$ 0.1 & 62 $\pm$ 1 & 34 $\pm$ 1 & 3.2 $\pm$ 0.2 \\ 
      SF & field & 53 $\pm$ 1  & 9.1 $\pm$ 0.2  & 2.4 $\pm$ 0.1 & 61.6 $\pm$ 0.6 & 33.4 $\pm$ 0.4 & 3.5 $\pm$ 0.1 \\
      AGN & void  &  25 $\pm$ 8  & 22 $\pm$ 7  & 45 $\pm$ 12 & 31 $\pm$ 9 & 65 $\pm$ 16 & 3 $\pm$ 2   \\ 
      AGN & wall  & 24 $\pm$ 3  & 23 $\pm$ 3  & 37 $\pm$ 4 & 26 $\pm$ 3 & 68 $\pm$ 6 & 5 $\pm$ 1  \\ 
      AGN & field &  25 $\pm$ 2  & 20 $\pm$ 1  & 37 $\pm$ 2  & 28 $\pm$ 2 & 67 $\pm$ 3 & 4.2 $\pm$ 0.5\\
      passive  & void  & 4 $\pm$ 1  & 3 $\pm$ 1  & 2 $\pm$ 1 & 5 $\pm$ 1 & 7 $\pm$ 2 & 34 $\pm$ 4  \\ 
      passive  & wall  &   2.4 $\pm$ 0.2  & 2.3 $\pm$ 0.2  &  2.6 $\pm$ 0.2 & 3.4 $\pm$ 0.3 & 5.5 $\pm$ 0.3 & 35 $\pm$  1 \\ 
      passive  & field &  2.3 $\pm$ 0.1  & 1.9 $\pm$ 0.1  & 2.5 $\pm$ 0.1 & 3.0 $\pm$ 0.1 & 4.9 $\pm$ 0.2 & 32.4 $\pm$ 0.5 \\
      \noalign{\vglue 0.2em}
            \noalign{\vglue 0.2em}
          \hline
      \hline
    \end{tabular}
    \label{table:diagcomp}
  \end{center}
\end{table*}

Since AGN emission is associated to accretion onto the massive central black hole, galaxy environment is expected to be correlated to AGN characteristics. This connection may be related to gas rich environments frequent in low and mildly high density regions and where galaxy interactions may play an important role as well.

Following this line, we have investigated possible differences in the classification of AGN comparing the three methods used in this work to select active objects. We also consider the different large-scale environments defined previously.
To accomplish this, we have examined the results of the different classification 
schemes addressed in this work taking into account the environment of the galaxies ie.
 voids, walls or field.
 We calculate the percentage of galaxies of a given class according to one classification criterium, 
 either BPT, WHAN or WISE, considering their classification using the other diagnostics tests, i.e. the percentage of cross--classes.
The results for void, wall and field regions are summarized in
Table \ref{table:diagcomp}.  
The table is divided into three parts where the BPT, WHAN and WISE comparison
are displayed in the upper, middle and lower parts respectively,
as it is indicated in the table. The errors correspond to a Poisson estimate in all cases.

For instance, the upper part of the table shows the comparison of BPT SF, composite and AGN galaxies (first column) with WHAN (columns 3-5) and WISE (columns 6-8) categories. In this table we show the percentages of cross--classes in the different diagnostic diagram.
The second column in the table indicates the large--scale region. 
Notice that the percentages do not add up to 100 due to rounding, and in the case of WISE galaxies there is 
a set of intermediate objects between SF and passive (1.5 $<(\rm w2-\rm w3)<$3.0) that we have not considered in this analysis.
The entries showing and horizontal bar indicate the lack of galaxies of the corresponding class in the large scale region.
The middle and lower parts of the table have a similar organization, for WHAN and WISE comparison, respectively. The content of the columns are indicated in their headings. 

As it can be seen in the table, there are no significant differences in the AGN fractions according to the large-scale environment in any of the cases analyzed.
This result suggests that the differences between the classification assigned to each galaxy are related to their criteria and would not be affected by the large-scale environment. 

\section{AGN abundances in void environment}

   \begin{figure}
   \epsfxsize=0.5\textwidth
      \centerline{\epsffile{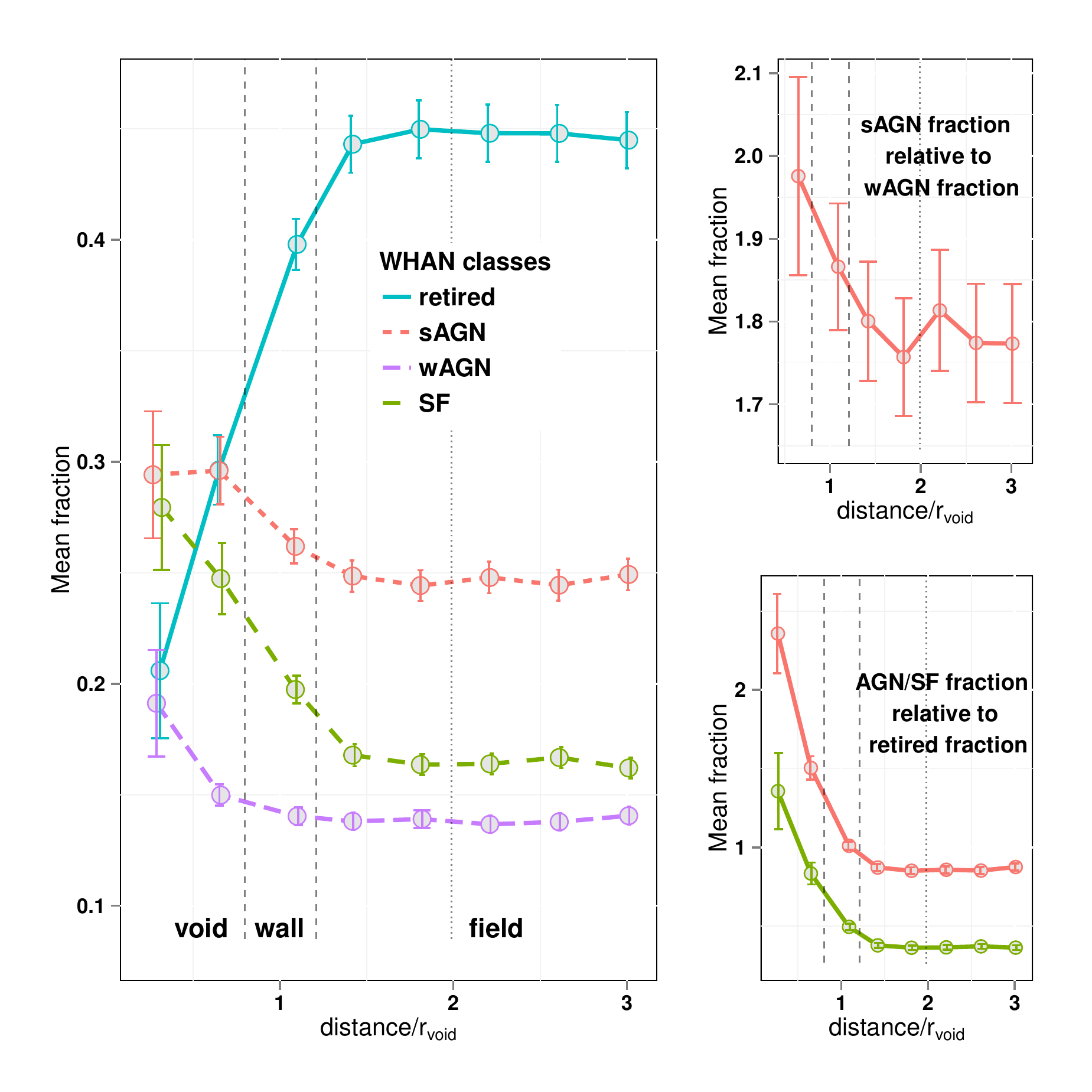}
   }
   \caption{
   {\it{Left panel:}} 
   Galaxy fractions for different WHAN classes as a function of void--centric distance.
   Colour and line types correspond to galaxy classes, as is indicated in the figure.  
   Large scale regions associated to void (field) are indicated by dashed (dotted) black lines.   
   {\it{Upper right panel:}} 
    sAGN fraction relative to wAGN fraction. 
   {\it{Lower right panel:}} 
    AGN (wAGN+sAGN) and SF galaxies fractions relative to retired fraction, in red and green lines respectively.
    }
   \label{fig:whanfrac}
   \end{figure}
%
   \begin{figure}
   \epsfxsize=0.5\textwidth
   \centerline{\epsffile{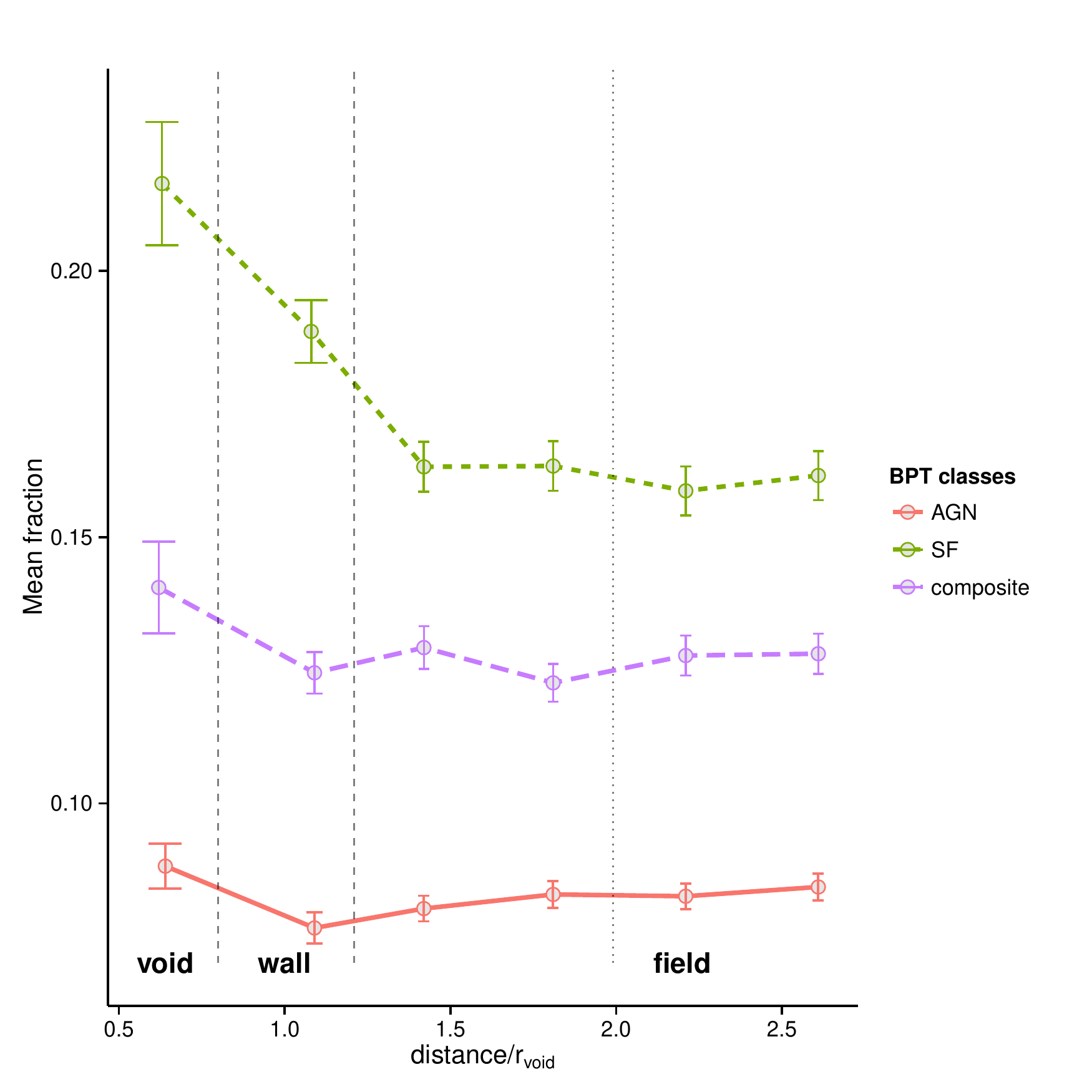}
   }
   \caption{
   Relative abundance of BPT AGN  (solid red line), star forming (dashed green line) and composite (long--dashed purple line) galaxies, as a function of void--centric  distance.
   Large scale region limits associated to voids (field) are given by dashed (dotted) black lines.   
    }
   \label{fig:fhlo3mhb}
   \end{figure}
%
   \begin{figure}
   \epsfxsize=0.5\textwidth
   \centerline{\epsffile{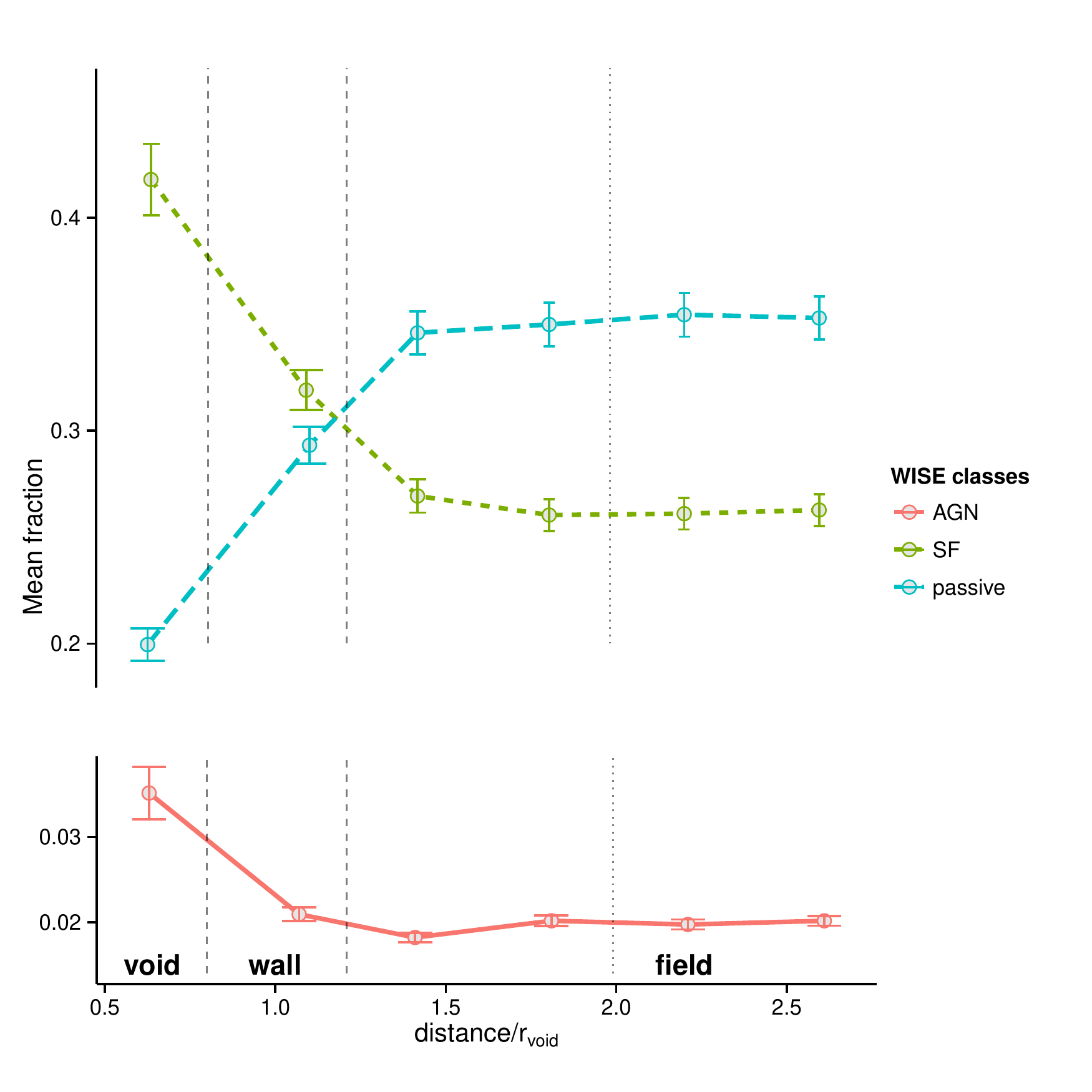}
   }
   \caption{
   Fractions of WISE AGN (solid red line), SF (dashed green line), and passive (long--dashed blue line) galaxies as a function of void--centric 
   distance. Large scale region limits associated to voids (field) are given by dashed (dotted) black lines.  
    }
   \label{fig:wisefrac}
   \end{figure}

Here we analyse the population of active objects in void, wall, and field regions by computing the mean fractions of different galaxy classes in these environments.

We calculate the mean galaxy class fraction ($f_{C}$) as a function of the 
distance to the void center (d) as a useful characterization of the void AGN connection,
\begin{equation}
f_{C}(r)=<N_{C}(r)/N_T(r)>, 
\end{equation}
where 
$N_{C}$ is the number of galaxies of a given class and 
$N_T$ is the total number of galaxies. 
We consider the average values for all identified voids in the sample.

The derived relative abundance of AGN according WHAN classification (see section \ref{sec:whan}) as a function of void radius are given in Fig. \ref{fig:whanfrac}, where dashed red and long--dashed purple lines correspond to sAGN and wAGN classes, long--dashed green to SF galaxies and solid blue to retired objects.
As it can be seen in the left panel of this figure, the fraction of sAGN significantly increases from walls to voids, consistent with a sharp decrease of the fraction of retired galaxies. Moreover, the fraction of sAGN, wAGN and SF galaxies are all higher in voids, and correspond to a decrease of approximately 60$\%$ of retired galaxies.
We also observe that the fraction of SF galaxies monotonically decreases as voidcentric distance increases, in contrast to retired galaxies whose fraction increases with distance. This behaviour is 
consistent with previous results of galaxy properties in voids
\citep{rojas_spec, hoyle_luminosity_2005,beygu2017}.
In the upper right panel of Fig. \ref{fig:whanfrac} we show the mean fraction of sAGN to wAGN galaxies.
A clear maximum in this relative fraction occurs at void interior and  walls. 
This result is consistent with an increase of the strength of activity in the nucleus in global low-density environments.
In the lower right panel of Fig. \ref{fig:whanfrac} we show the mean relative fraction of AGN (sAGN+wAGN) to retired galaxies in red lines.
As can be seen in the figure, the increment of the relative fraction of AGN to retired as a function of voidcentric distance indicates that AGN galaxies dominate in void interiors.
We also show the mean relative fraction of SF to retired galaxies in green lines, which grows towards the voids, as expected according to the well-known properties of galaxies in these environments.

In a similar fashion, we have also explored the relative abundance of galaxies classified according to the BPT diagram as a function of the distance to the void centres. 
The results are shown in figure \ref{fig:fhlo3mhb} where each colour and line type indicate a class of galaxies; solid red, dashed green, and long--dashed purple lines correspond to AGN, SF, and composite galaxies, respectively. Given the lower number of objects in this sample only one bin was taken in void interiors.
The vertical dot lines mark the boundaries between the large-scale low-density regions, namely voids and walls, with the field, and are also indicated in the figure. 
As can be seen in the figure, the fractions of the three classes of galaxies analyzed increase in voids, this result is consistent with those obtained for the SF and AGN galaxies according to the WHAN classification (dashed lines in Fig. \ref{fig:whanfrac}).

The fact that WISE AGN may include a significant fraction of obscured AGN could make them suitable to explore environments rich in gas and dust. Accordingly, we have also analysed the relative abundance of galaxies classified according to WISE colors using our environment definition. The resulting fractions of these classes of galaxies as a function of distance to void centres are displayed in Fig. \ref{fig:wisefrac}.
As in Fig. \ref{fig:fhlo3mhb}, only one bin was taken in void interiors, where it can be seen a significant increase in the fraction of WISE AGN (solid red line) and SF (dashed green line), while a notable decrease is observed in the fraction of passive galaxies in voids (long--dashed blue line).

The results reported in this section show strong evidence of an increase of the relative population of active galaxies and AGN emission power in voids, this result is independent of the AGN classification scheme for 3 different diagnostic diagrams (figures \ref{fig:whanfrac}, \ref{fig:fhlo3mhb} and \ref{fig:wisefrac}). 
We want to emphasize that these findings are entirely consistent with the already accepted results regarding the preponderance of galaxies with star--forming activity (figures \ref{fig:whanfrac}, \ref{fig:fhlo3mhb} and \ref{fig:wisefrac}) and the low population of passive galaxies (figures \ref{fig:whanfrac} and \ref{fig:wisefrac}) in voids. 

%

\begin{figure*}
\centering
\subfloat[Mean values of the EW(H$\alpha$) as a function to void centric distance, for SF (solid green line), sAGN (dashed red line) 
and wAGN (long--dashed purple line) galaxies accordingly WHAN criteria.   
   As in previous figures, large scale regions associated to voids (field) are indicated by black dashed (dotted) lines. 
\label{fig:meanprop}]{\includegraphics[width=0.49\textwidth]{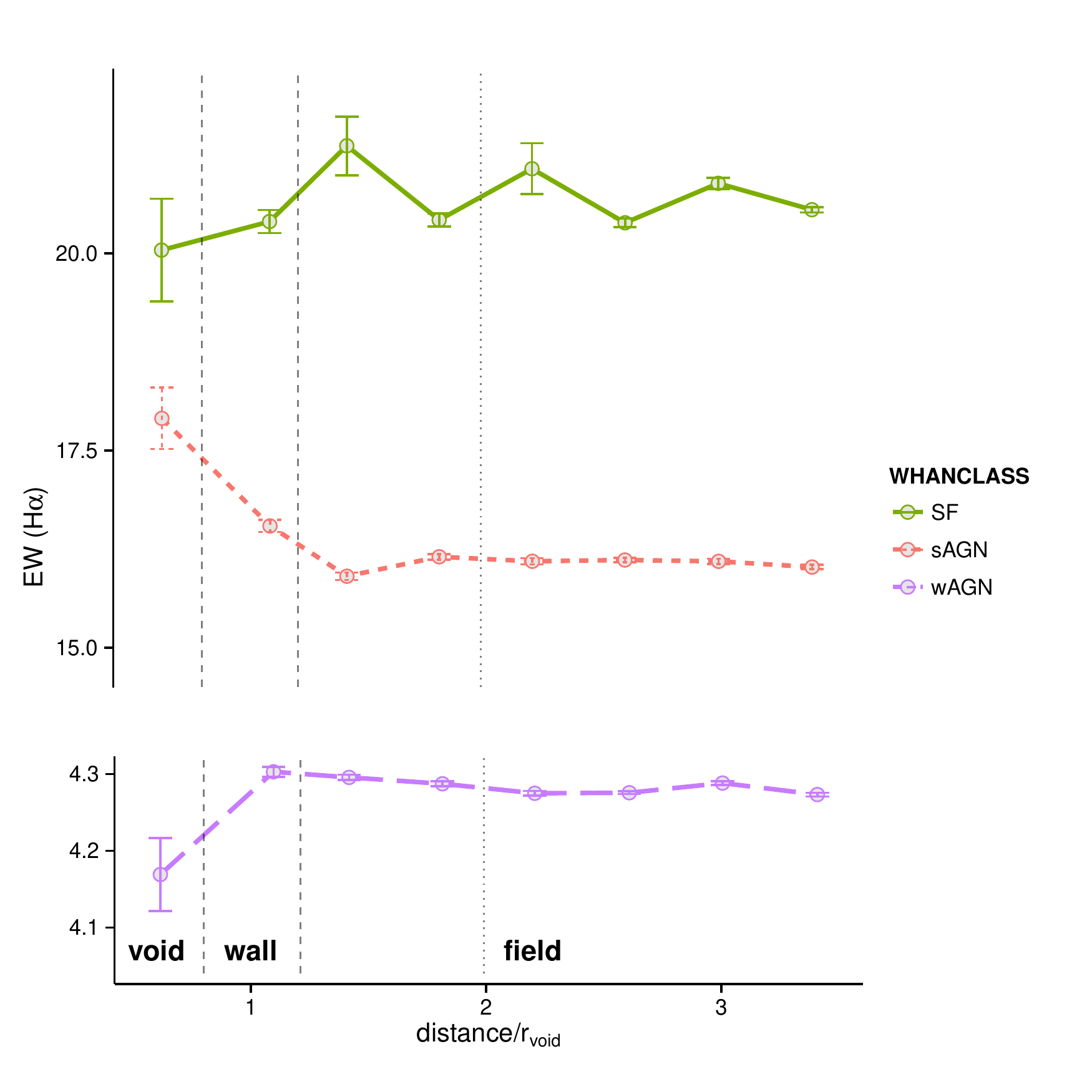}}\hfill
\subfloat[Density distribution of  EW(H${\alpha}$) for WHAN strong AGN (bottom panel) and star 
        forming galaxies (upper panel) at voids walls and field in SDSS sample. 
        \label{fig:densewha}]{\includegraphics[width=0.49\textwidth]{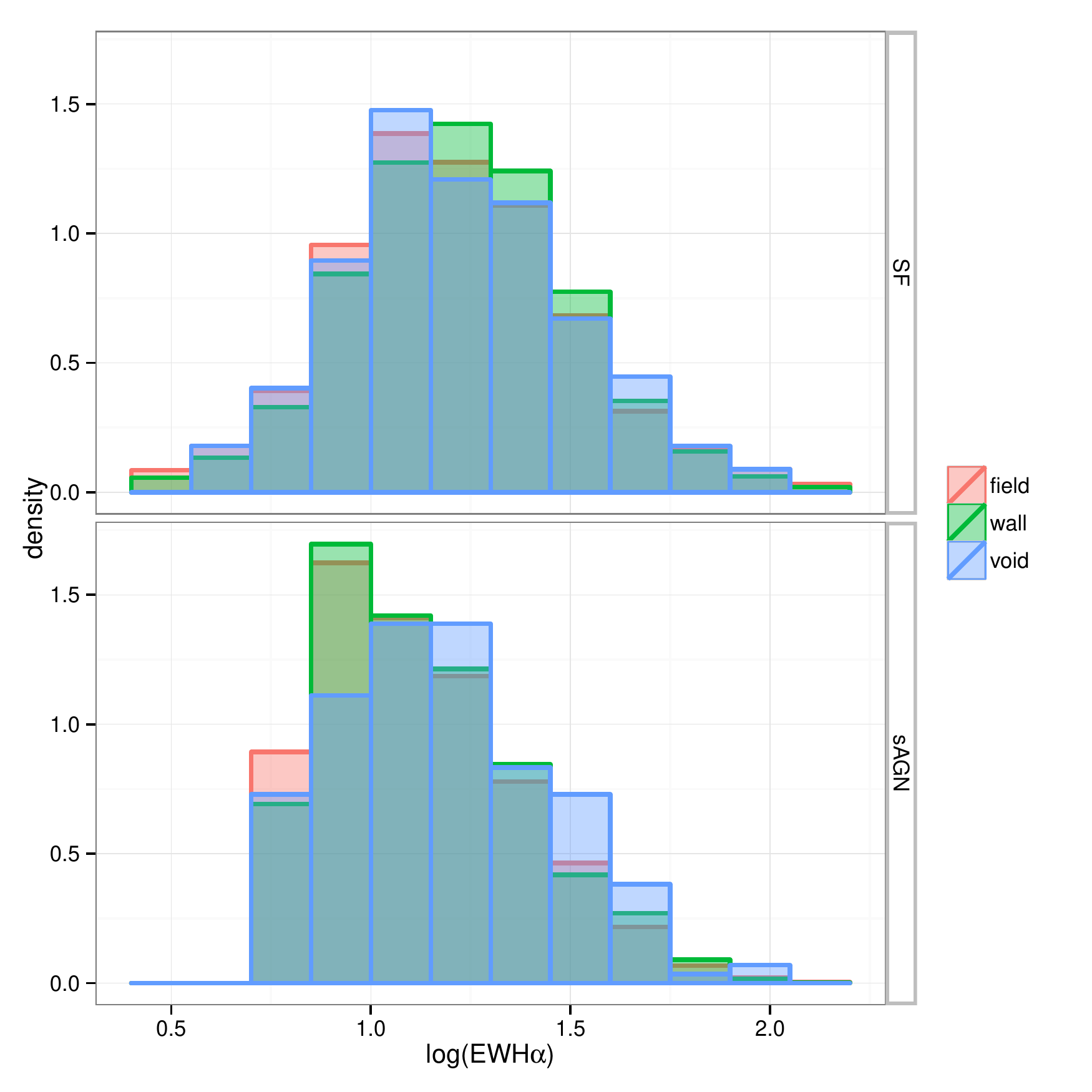}}\hfill
\caption{Study of the EW(H$\alpha$) for the different WHAN classes.} \label{fig:12}
\end{figure*}
%
\section{AGN activity around voids}
In this section we explore the AGN energy release as inferred by the [OIII] emission line luminosity
(L[OIII]). Also, we analyse EW(H$\alpha$) line values and black hole mass proxies from correlations with the central stellar velocity dispersion. As in previous sections, 
we have considered separately, void, wall and field regions.
\subsection{The equivalent width of H$\alpha$}
H$\alpha$ emission is related to the number of massive
stars and is a direct measurement of the star formation rate. 
Nevertheless, besides its dependence on star formation activity, the H$\alpha$ luminosity is also affected by AGN activity. For instance, values of EW(H$\alpha$) are used in WHAN diagrams to separate strong from weak AGN. Moreover, it is also well known that Syfert galaxies have higher values of EW(H$\alpha$) than LINERs \citep[][]{cidfernandes2010}.
Taking into account these issues, we have examined the mean EW(H$\alpha$) values as a function of the distance to voids for both, active and star--forming 
galaxies classified according to the WHAN diagnostic diagram. The results are 
shown in Fig. \ref{fig:meanprop}. 
As it can be seen in this figure, EW(H$\alpha$) values increase in voids for sAGN
(dashed red line) while on the contrary, they decrease for wAGN (long--dashed purple line). On the other hand,
for SF galaxies, no significant dependence are evident (solid green 
line). \\
We have also explored the EW(H$\alpha$) distribution of sAGN and SF 
galaxies 
in voids, walls and field regions, and show the results in Fig. \ref{fig:densewha}.
The upper panel corresponds to SF galaxies, and the lower panel to sAGN. The line
colours correspond to our different definition of large scale environment (blue, green and red: void, shell and
field, respectively).
As it can be seen in this figure, the distribution of EW(H$\alpha$) values for sAGN galaxies in voids
differs substantially from that corresponding to wall and field environments (lower panel). This adds  
to the observed variation of the mean EW(H$\alpha$) values with voidcentric distance 
seen in the gray line of Fig. \ref{fig:meanprop}. 
On the other hand, the distribution and mean values of EW(H$\alpha$) for SF galaxies show no significant differences (upper panel of the Fig. \ref{fig:densewha}). 
We used the Kolmogorov--Smirnov (KS) test to 
evaluate the statistical significance of the differences between EW(H$\alpha$) distributions for different large scale environments. If the p--value is less than 0.05 we consider that the differences between distributions are highly significant. Since p-values obtained for SF galaxies are larger than 0.6, according to this test, the difference between the samples is not significant enough to confirm that they have different distribution.
On the other hand, we obtain p--values $<$ 0.01 for sAGN samples in voids and field, and in voids and walls. This confirms that the differences in the EW(H$\alpha$) distributions are significant, whereas for wall and field sAGN the resulting p--values is greater than 0.01 indicating that the two samples come from the same distribution.
\\

The median (mean) logarithmic values of the EW(H$\alpha$) distributions of SF and sAGN galaxies in voids yield 
values 1.18 (1.22) and 1.15 (1.19) respectively.
These results show evidence of higher EW(H$\alpha$) values for star--forming galaxies than for AGN in void environment,
consistent with the overall larger star formation activity characterizing these regions.
However, if we compare to field galaxies, we obtain that the ratio
between the medians of the log EW(H$\alpha$) distributions for sAGN and SF galaxies are 0.062 and 0.013 respectively. This result indicates that the H$\alpha$ equivalent widths of sAGN are significantly more affected by the void large-scale environment.

In WHAN star-forming galaxies, although their relative population in voids increases 
(long--dashed green line in Fig. \ref{fig:whanfrac}), 
their H$\alpha$ equivalent widths remain approximately constant in different large--scale regions  (solid green line in
Fig. \ref{fig:meanprop}). Nevertheless, 
we find higher fractions and a noticeable increment of  H$\alpha$ equivalent widths for WHAN sAGN 
(dashed red lines in Figs. \ref{fig:whanfrac} and \ref{fig:meanprop}), 
consistent to an increment of AGN activity in voids.
This behavior, which seems to be characteristic of active galaxies, could account for the effects of the global environment defined by voids in destabilizing the central regions of the galaxies.


\subsection{The [OIII] luminosity}

The [OIII] luminosity (L[OIII]) is considered a suitable proxy of the AGN activity.
Although this line is also suitable to trace for the presence of massive stars, it is relatively 
weak in star--forming galaxies which are rich in metals \citep[][]{kauffmann2003,Heckman2004}.
An advantage of using the [OIII] line is its strength and the fact that it can be easily detected in most galaxies. In this work, we use the [OIII] line corrected for optical reddening using the Balmer decrement and the obscuration curve of \citet{Calzetti2000}.

\begin{figure*}
\centering
\subfloat[BPT AGN mean $\rm L \lbrack \rm O \rm III\rbrack$ values as a function of voidcentric distance. The colour intensities correspond to different subsamples selected according to black hole mass distribution percentiles. The black hole mass thresholds and the corresponding percentiles are $\rm log(M_{\rm BH}/M_{\odot}) \geq $ 7.29, 7.64, 7.95, 8.15 and 8.45, associated to the 25, 50, 75, 80 and 90 percentiles respectively. Large scale regions associated to voids (field) are indicated by dashed (dotted) black lines.  \label{fig:meanlo3}]{\includegraphics[width=0.49\textwidth]{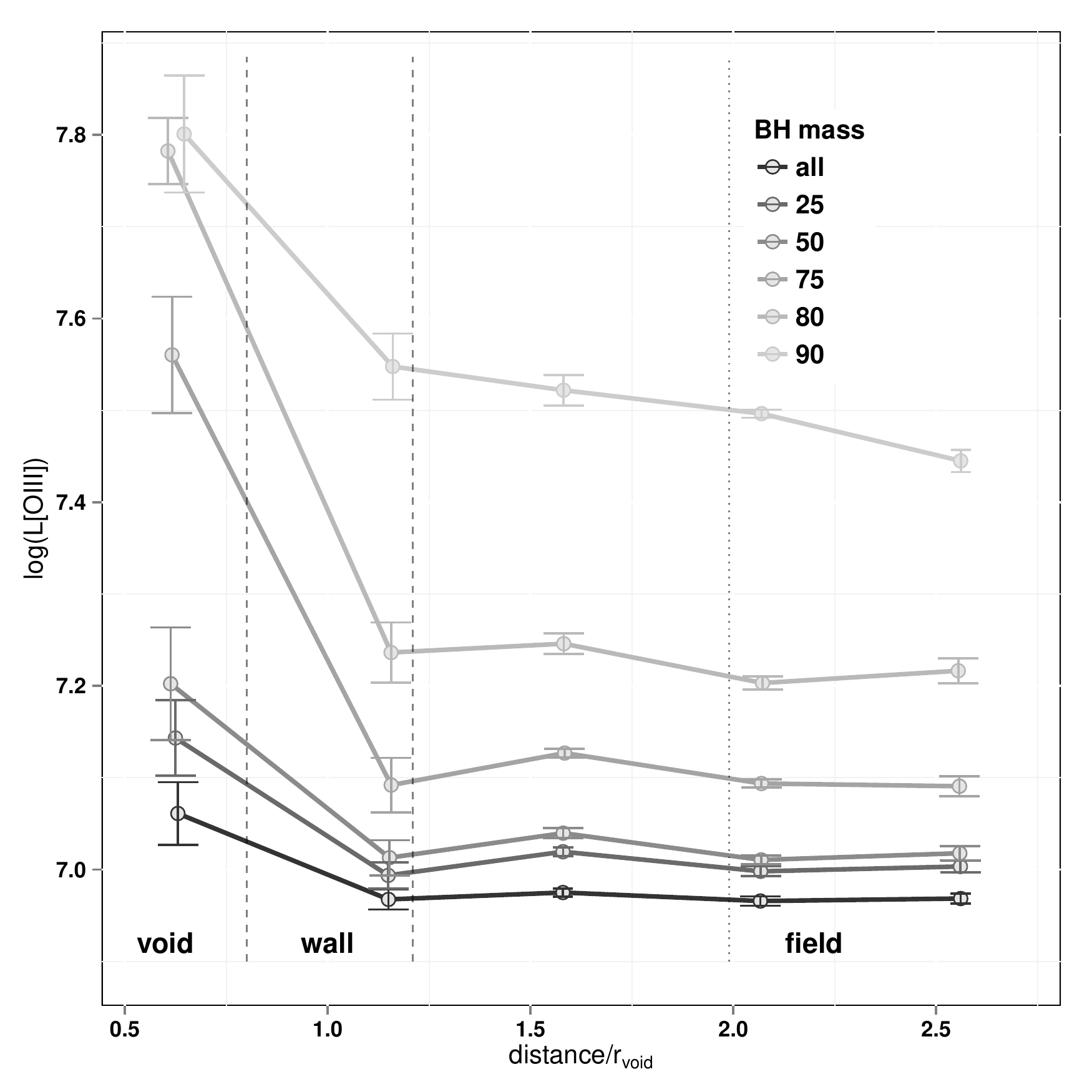}}\hfill 
\subfloat[{\it{Upper panel:}} Normalized distribution of the $\lbrack \rm O \rm III\rbrack$ luminosity for the full samples of BPT AGN in voids (blue), walls (green) and field (red).  {\it{Lower panels:}} As upper panel but for subsamples selected according to different percentiles of the central black hole mass distribution (black hole mass percentiles are indicated in the upper labels and correspond to the same values than in Fig. \ref{fig:meanlo3} ). \label{fig:denslo3_2}]{\includegraphics[width=0.49\textwidth]{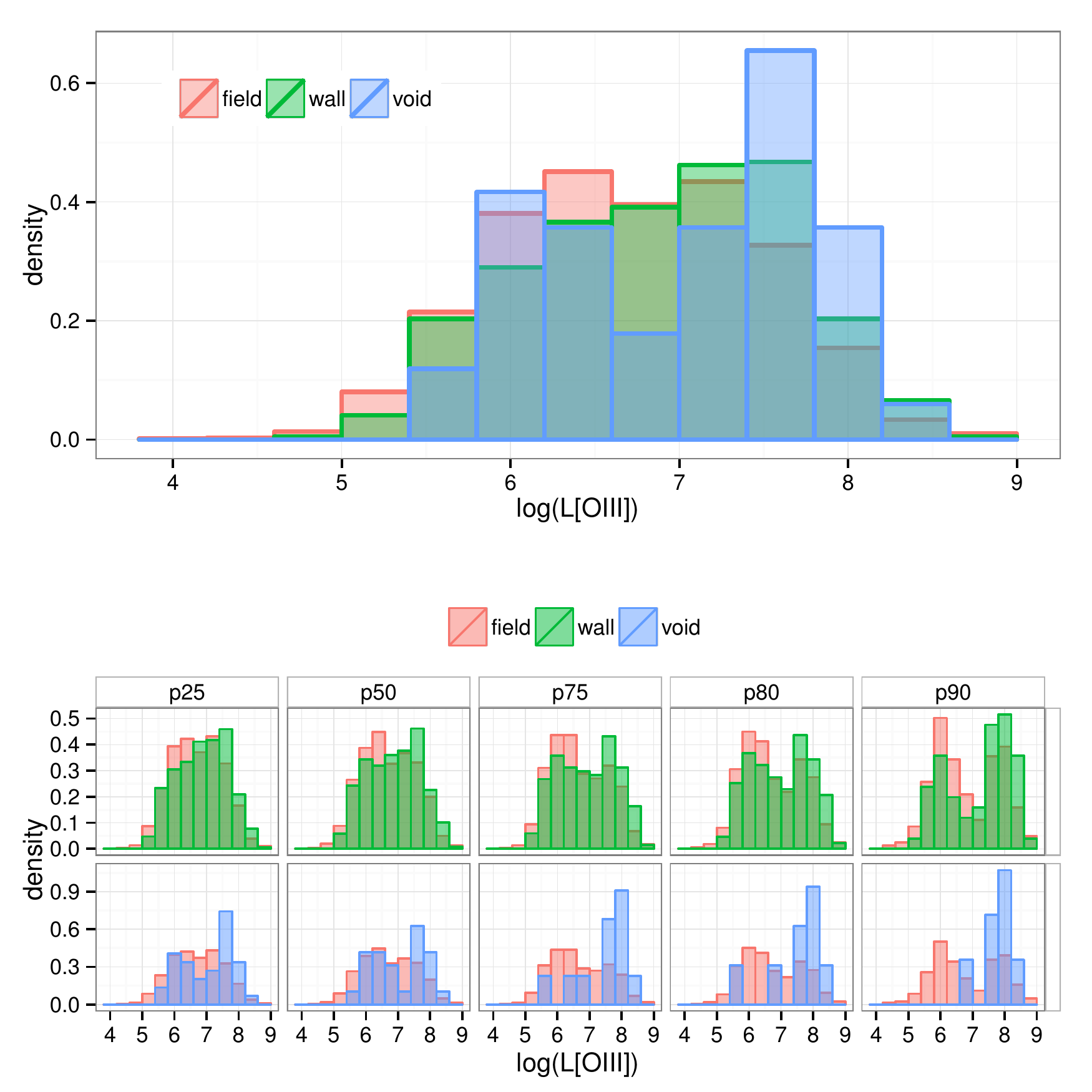}}\hfill
\caption{Study of the [OIII] luminosity for the BPT AGN.} \label{fig:mlo3}
\end{figure*}

We study  the [OIII] luminosity  of BPT AGN populating void, wall and field regions. To analyze this we examine the distribution of L[OIII] for AGN galaxies in voids, walls and field that is represented in the upper panel of the Fig. \ref{fig:denslo3_2}. Each color corresponds to a different large-scale region as indicated in the figure. As can be seen, the distributions are remarkably different. The statistical significance of these differences is reinforced with the p-values obtained from the KS test, which are less than 0.05.

We also consider L[OIII] dependence on central black hole mass. To this end, we use the observed correlation between black hole mass, $\rm M_{\rm BH}$, and  bulge velocity dispersion, $\sigma^*$ as given by 
\citet{tremaine2002}. Thus, we adopt as a proxy of black hole (BH) mass the relation
$$\rm log(M_{\rm BH}/M_{\odot}) = 8.13 + 4.2 log(\sigma^*/200 km s^{-1})$$ Given the instrumental resolution of SDSS spectra, $\sigma^*\sim 60-70 \rm kms^{-1}$, for this analysis we have filtered out galaxies considering $\sigma^*> 70 \rm kms^{-1}$. 

We have considered subsamples selected according to percentiles
in the distribution of black hole mass proxy, and in Fig. \ref{fig:denslo3_2} we show the distribution of L[OIII] for BPT AGN in void, wall and field environments with BH mass greater than percentiles 25, 50, 75, 80 and 90 (ie. log(M$_{\rm BH}$/$\rm M_\odot$)$\geq$ 7.29, 7.64, 8.03, 8.15 and 8.45, respectively.).
The upper panels provide the results of wall regions (green), and the lower panels void (blue) environments. In both cases the field region (red) is shown for comparison. Black hole mass percentiles are indicated in the upper labels. As it can be seen, for all cases, AGN in voids and walls exhibit larger L[OIII] values than their field counterparts.

In order to study the dependence of the [OIII] luminosity on voidcentric distance, we show in Fig. \ref{fig:meanlo3} the mean L[OIII] values for different lower thresholds of BH mass proxy (those associated to the previously adopted BH mass percentiles namely  $\rm log(M_{\rm BH}/M_{\odot}) \geq$ 7.29, 7.64, 8.03, 8.15 and 8.45, corresponding to the 25, 50, 75, 80 and 90 percentiles respectively). In this figure, these BH mass lower limits are shown in different colour intensity, and the darkest one represents the full AGN sample. From this figure it can be seen that AGN in void regions exhibit the largest L[OIII] values whereas the smaller ones are found in the field. Moreover, there is a significant increase in L[OIII] mean values in void regions as reflected in the higher values as the distance to the void decreases. This is observed for all BH mass thresholds although it is more noticeable for the most massive BHs (p75, p80, 90, the 3 lighter lines in the figure). We notice that due to the low number of AGN in voids, we have only considered one bin within the void radius.

\begin{figure}
   \epsfxsize=0.5\textwidth
   \centerline{\epsffile{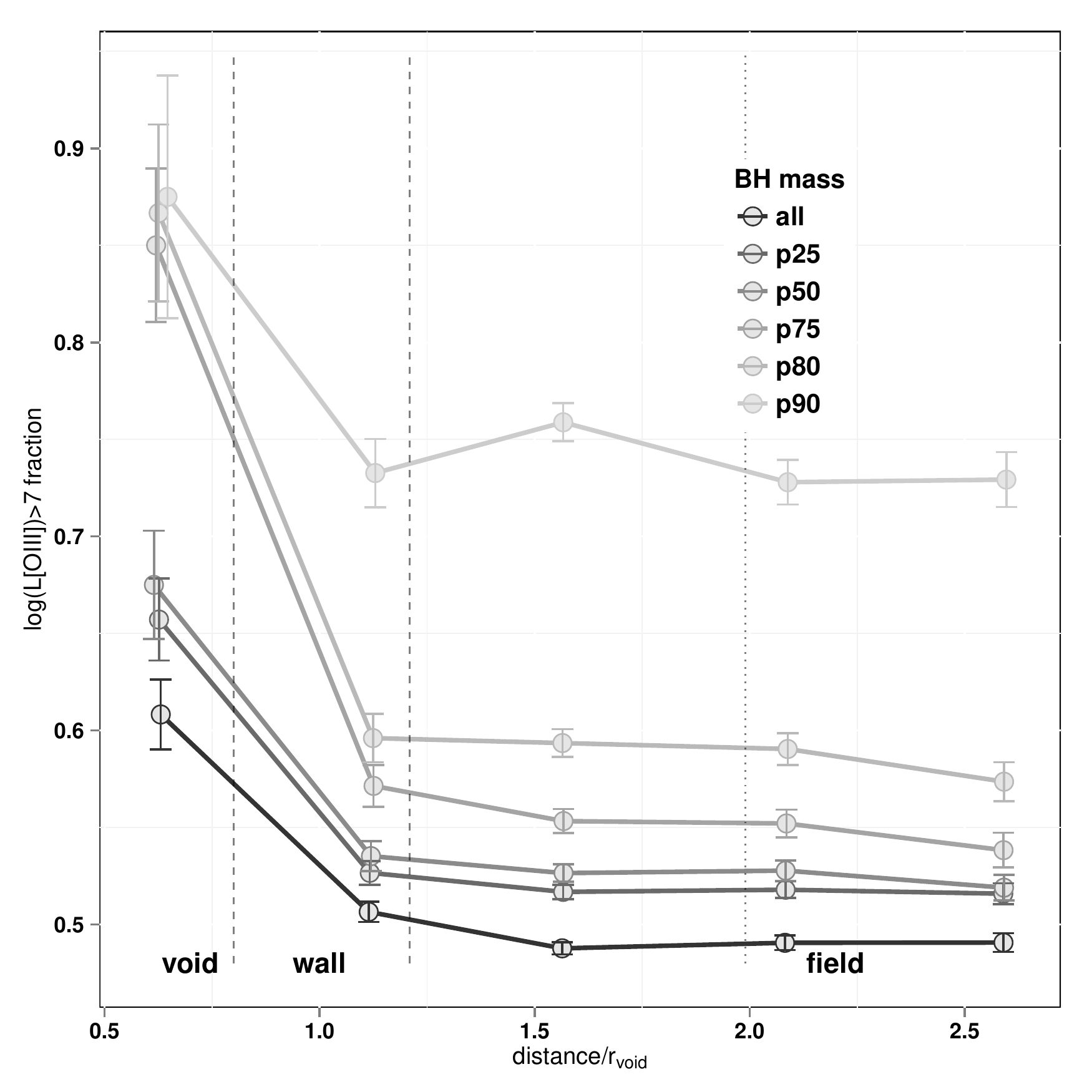}
   }
   \caption{
   Relative abundance of BPT AGN with high [OIII] luminosity as a function of voidcentric distance for different black hole mass thresholds ($\rm log(M_{\rm BH}/M_{\odot})\geq $ 7.29, 7.64, 8.03, 8.15 and 8.45, associated to the 25, 50, 75, 80 and 90 percentiles respectively). Black hole mass increases for lighter colours as indicated in the figure. Large scale regions associated to voids (field) are indicated by dashed (dotted) black lines.
    }
  \label{fig:fhlo3}
 \end{figure}

Since according to WHAN classification, sAGN dominate over wAGN in voids (see right upper panel of Fig. \ref{fig:whanfrac}), we have selected BPT AGN with high [OIII] luminosity to further explore this effect. 
Our procedure follows the use of L[OIII] as a tracer of the strength of nuclear activity and
adopt the criterium that the most powerful AGN are those with L[OIII] $>$ 10$^7$ L${\odot}$. A similar limit has been adopted previously by several authors to select powerful AGN  \citep[e.g.][]{kauffmann2003,Coldwell2009,Alonso2013,Duplancic2021}.
We study the population of powerful AGN by analysing their relative fraction as a function of voidcentric distance. The results are portrayed in Fig. \ref{fig:fhlo3}, where each colour corresponds to different lower limits of BH mass associated to the same percentiles as in Fig. \ref{fig:meanlo3}.
Similarly, the result for the full AGN sample is shown in the darkest line of Fig. \ref{fig:fhlo3}, and we only considered one bin within the void radius due to the low number of AGN in voids in each subsample.
As it can be seen in this figure, there is a relevant population of powerful AGN in void regions since the fraction of high L[OIII] values increases towards void centres. Moreover, this trend is more noticeable for the most massive BH subsamples (percentiles 75, 80, 90). \\

   \begin{figure}
   \epsfxsize=0.5\textwidth
   \centerline{\epsffile{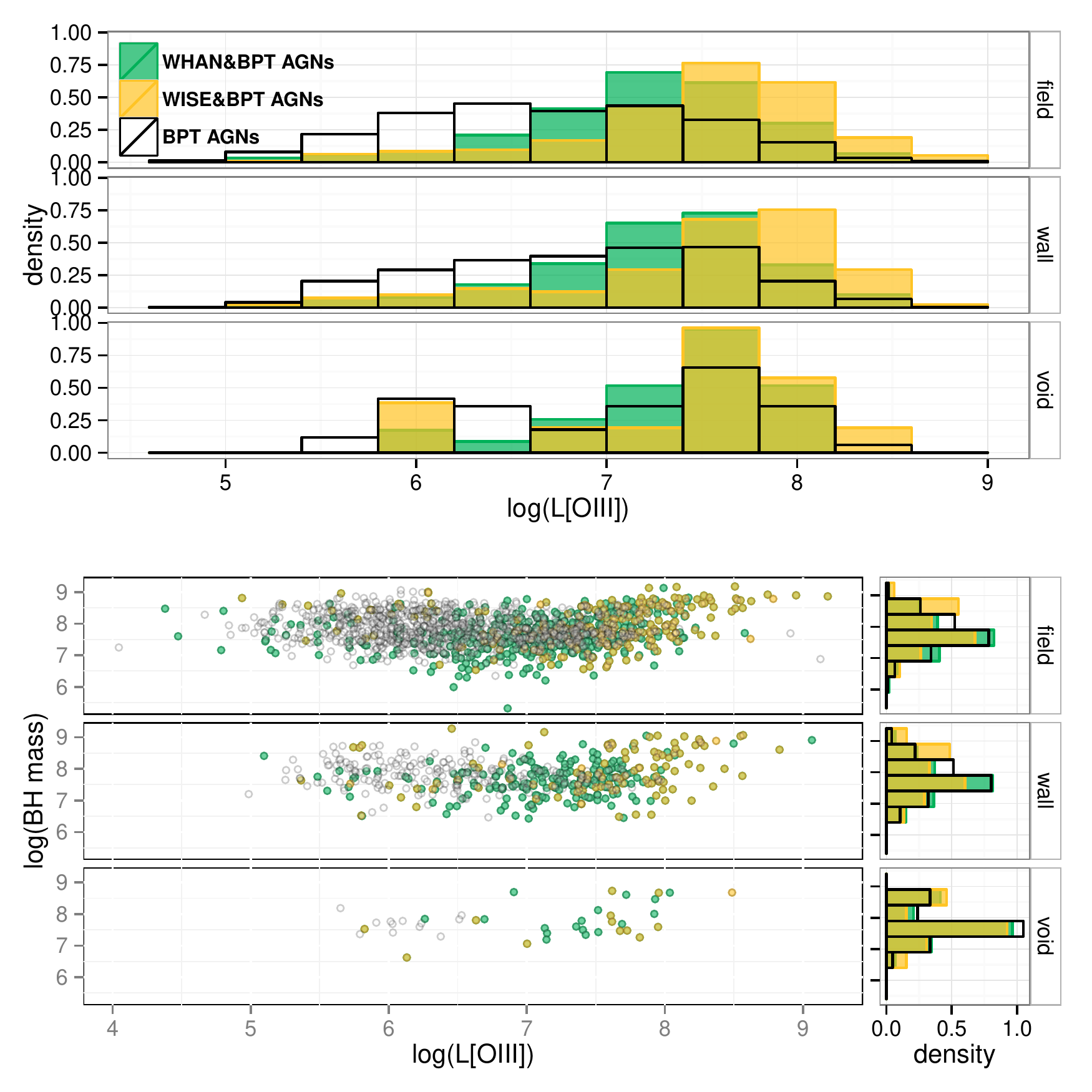}
   }
   \caption{
   {\it{Upper panel:}}
   Normalized distribution of the [OIII] luminosity for BPT AGN which are also classified as AGN following  
    WISE (yellow) and WHAN (green) criteria. 
    Black lines shows the L[OIII] distribution for BPT AGN. 
    Each panel indicates galaxies in void, wall and field regions.
   {\it{Bottom panel:}}
      Scatter plots of the black hole mass estimates vs L[OIII] for AGN in voids, walls and field, with marginal distributions for black hole mass estimates. 
   The different colours correspond to AGN selection criteria as in the upper panel.
    }
   \label{fig:denslo3}
   \end{figure}

\begin{figure}
   \epsfxsize=0.5\textwidth
   \centerline{\epsffile{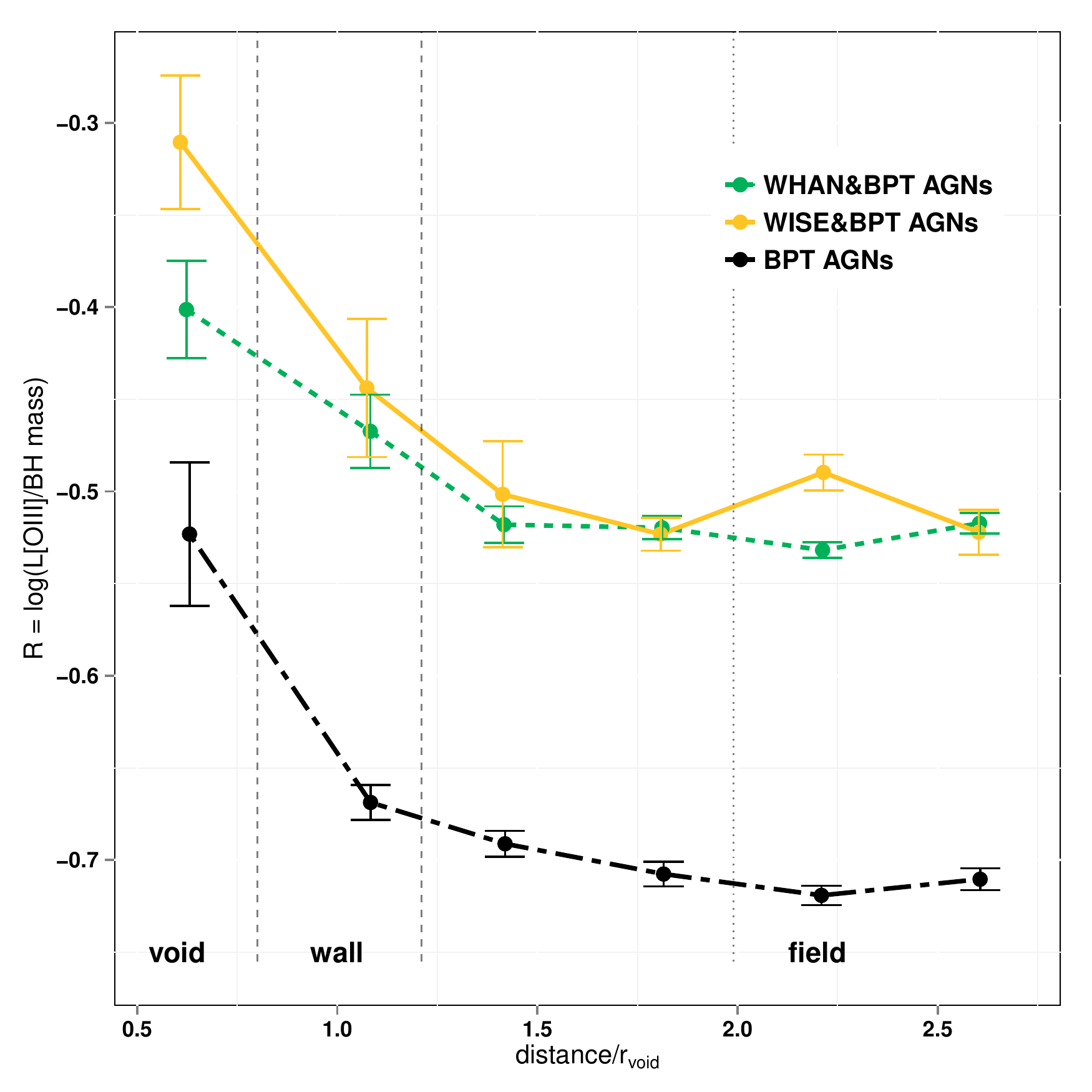}
   }
   \caption{
   AGN efficiency in void, wall and field, for galaxies satisfying the AGN classification criteria of WHAN and BPT (dashed green line), WISE and WHAN (solid yellow line) and BPT (two--dashed black line).
        }
  \label{fig:efficiencia}
 \end{figure}

We have also explored samples of AGN selected in the infrared provided by WISE.  Since several Optical BPT AGN are also identified as mid--IR active objects by using WISE colour--colour diagram,
here we explore what is the role of these galaxies in void regions and analyse their influence in the peak of the L[OIII] distribution. 
The resulting distributions are displayed in the upper panel of Fig. \ref{fig:denslo3}, where we show the L[OIII] distribution of AGN selected by both WISE and BPT criteria (yellow).
As it can be seen, the distributions look similar in void, wall and field, their median values are indistinguishable as well (p-values from KS test are $>$ 0.2 for all combination of subsamples).  
We also display the L[OIII] of BPT AGN in black lines (notice that these distributions are already shown in the in upper panel of Fig. \ref{fig:denslo3_2} and are displayed here to facilitate comparisons). 
As it can be noticed, AGN WISE selection is preferentially associated to powerful optical AGN in all environments examined here
(aproximatelly 80\% have L[OIII]$>10^7\ L_{\odot}$).\\
Additionally, for comparison we have explored the contribution of WHAN sAGN to the L[OIII] distribution in voids, walls and field. We consider sAGN that are also classified as active objects by BPT. The results are displayed in green in the upper panel of Fig.  \ref{fig:denslo3}. As expected WHAN sAGN are also associates to powerful AGN selected according to their [OIII] luminosity (the percentage of galaxies with  L[OIII]$>10^7\ L_{\odot}$ ranges from 70 to 80\% when going from field to void). 
As can be easily seen in the figure, WISE AGN are the most powerful objects in general, regardless of the global environments analyzed here.
Nonetheless, when we examine the galaxies in voids, we obtain a remarkable fraction of powerful AGN, whose value (L[OIII]$>10^7\ L_{\odot}$ fraction $\sim$ 0.8 ) is similar for any of the criteria adopted to select them. 
Furthermore, the [OIII] luminosity distributions of the AGN subsamples presented in the figure are comparable both in means and in dispersion when they are in voids, unlike what we obtain when these galaxies reside in walls and field. 
\\
We have also analyzed the relations between BH mass estimates and [OIII] luminosity for AGN in voids, walls and field, which are displayed in scatter plots in the lower panels of Fig. \ref{fig:denslo3}.
We have studied subsamples of AGN selected according to the BPT criteria (black points), BTP that are also WISE AGN (yellow points) and WHAN sAGN also active according to the BPT diagnostic diagram (green points),  similarly to what was done in the upper panel of this figure. 
We also show the corresponding marginal distributions of BH mass estimates
in the vertical right panels.
From this figure it can be appreciated a slight correlation between BH mass and L[OIII] for WISE and WHAN subsamples (yellow and green points), moreover WISE AGN seems to be associated not only to the most powerful objects but also to galaxies with larger BH mass, especially for wall and field galaxies.
\\
Based on the previous results it is interesting to study the accretion strength of the central black holes. To this end we calculate the accretion rate parameter, $\cal R=\rm log(\rm L[OIII]/M_{\rm BH})$ \citep[][]{Heckman2004}
that provides a widely used estimate of the AGN efficiency.
 We explore this efficiency parameter as a function of voidcentric distance for the AGN subsamples described in Fig. \ref{fig:denslo3} and
show the results in Fig. \ref{fig:efficiencia}.
From  this  analysis  we  show that WISE AGN and WHAN sAGN exhibit a higher fraction of efficient accretion rate with respect to the BPT AGN. This trend is independent of the large--scale environment. Moreover there is an increment in the mean acretion rate with decreasing voidentric distance and in void interiors WISE AGN display the higher $\cal R$ values in comparison with active galaxies selected acordin to BPT and WHAN.

\citet{Alonso2018} study the influence of bars and interaction on AGN properties. These authors select galaxies with high accretion rate considering $\cal R$ $>$ -0.6 and found that barred AGN galaxies show a moderate excess of efficient accretion rate with respect to active galaxies in pairs and in a control sample. In this work we find that within voids AGN have $\cal R$ mean values well above this threshold. Moreover all BPT active galaxies also selected as AGN by WISE or as sAGN by WHAN are high accreting objects. 
%

\section{Summary and Results}
In the present work we identify star forming, AGN and passive galaxies in a SDSS volume limited galaxy sample and study the influence of large--scale environment by considering the distance to cosmic voids centres. We use BPT and WHAN optical line diagnostic diagrams as well as mid--IR color--color diagrams based on WISE data.
 By following this approach we can perform a detailed analysis of the physical characteristics of active galaxies in an homogeneously identified samples of voids.

The main results of this work can be summarized in the following items:
\begin{itemize}
    \item An increment of both star--forming and AGN galaxies from walls toward void interiors, compared with field regions. These trends are independent of the classification schema, and are evident for WHAN, BPT and WISE star--forming and active objects identification criteria. 
   \item Decreasing fraction of WHAN and WISE passive objects from field toward void interiors. This is consistent with a lack of passive galaxies in underdense regions.   
   \item An increment of the H$\alpha$ equivalent widths of WHAN sAGN toward void regions consistent with more powerful objects associated to void environments.
   \item We find that BPT AGN in void regions show higher L[OIII] values for similar black hole mass ranges  with a tendency for [OIII] luminosities to increase with $\rm M_{\rm BH}$. Therefore, we expect AGN residing in voids to have larger [OIII] luminosity than counterparts with similar black hole masses located in a global density environment.
   \item  WISE AGN also selected as optical active objects have the highest [OIII] luminosities regardless their large-scale environment. Moreover, these AGN correspond to those with the highest accretion rate ($\cal R >$-0.6) in our analysis. These effects become stronger within the void environment.
 \end{itemize}

We further notice that examining these topics could help to better understand the role of the different accretion mechanisms that trigger nuclear activity in galaxies.
Overall, AGN feeding mechanisms are expected to be effective   
in local underdense regions which are rich in gas, part of which can 
be efficently conducted to the central regions of galaxies via radiative cooling. This mechanism could become a suitable source for accretion onto the center of galaxies and feed  galactic nuclei 
\citep{davies1973, solanes2001, diserego2007, grossi2009, cortese2011, catinella2013}. 
This scenario could be key to explain our findings of a larger fraction of AGN in voids and walls which are gas rich, locally underdense regions, with additional fuel by gas expanding from void interiors.

\section{Discussion and Conslusions}

The void underdense environment provides a simpler dynamical behavior compared to that of groups and clusters. For this reason, is is relevant to explore the possible effect of these environments on galaxy evolution, particularly star formation and central AGN accretion. 
In this context, nuclear activity in galaxy hosts residing in voids and surrounding regions can be important to understand the interplay between star formation and black hole accretion and the role of environment.   
The presence of AGN in cosmic voids had been reported by different authors \citep[][]{constantin2008,liu2015,Argudo2018}. Nevertheless the influence of large--scale environment on AGN activity is still on debate. \citet{Amiri2019} used volume limited samples of galaxies to compare the fraction of AGN in rich galaxy clusters and  voids finding that in the local Universe ($0.01<z<0.04$) AGN activity has a larger dependence on host stellar mass than on the local galaxy density. In the same line \citet{Miraghaei2020} also find small effects of environment on AGN activity although it is reported a higher fraction of optical AGN in massive red galaxies in voids compared to similar galaxies in dense environments. On the theoretical side, using cosmological hydrodynamical simulations \citet{Habouzit2020} find a coeval evolution of galaxies in voids and their central black holes, where an approximately 20 \% of them become active galactic nuclei.

In the present work we have study the relative fraction of galaxies with different activity classes as a function of distance to cosmic voids centres.
In order to analyse the activity of galaxies we use the BPT, WHAN and WISE
diagrams in a volume limited sample of galaxies extracted from the SDSS. 
Briefly, our results show a predominance of actively star--forming galaxies and an clear excess of AGN galaxies inside voids and in void surroundings. This result is independent of the classification scheme addressed (either BPT, WHAN or WISE).
The increment of the fraction of AGN and star--forming galaxies is evident from walls to void interiors. We also notice that although void walls correspond to the boundaries of cosmic voids, given our relatively strict void definition, these edges are regions still largely underdense in global expansion with respect to void centres. 
In this line, we recall previous studies of the effect of the large scale environment in walls surrounding voids as reported by \cite{ceccarelli_large-scale_2008,ceccarelli_low_2012}. 
 
We detected a significant deficit of retired and passive galaxies in voids when we studied the WHAN and WISE classifications. This behavior is consistent with the expected properties of galaxies in low--density regions.
This result, combined with the increase of the population of star--forming galaxies in voids, constitutes a further confirmation of the well-studied properties of void galaxies. 

A relevant issue concerning our results on AGN activity and the location of hosts with respect to void centres is the tendency of AGN galaxies to populate preferentially voids and walls rather than the field regions. This could be related to the fact that hosts residing these environments may have their central black hole accretion rate fuelled by gas flowing from void interior outwards the void. On the other hand in high density regions as the centres of galaxy clusters or rich groups it has been show a lack of AGN excess  \citep[e.g.][]{Popesso2006,Coldwell2014,Coldwell2017,Amiri2019}. This may be a consequence of low merging rate given the high velocity dispersion of these systems and also the interaction with cluster hot gas that may cause a  depletion of gas from galaxies.
Note that the global low--density environment is uniquely characterized by voids, while active galaxies have different selection criteria.
It is, therefore, interesting to examine whether the environment is reflected in different ways with different selection criteria.
To analyze this, we use 3 different schemes to classify and select AGNs and we obtain consistent results with all of them, giving extra strength to the results. It is worth mentioning that although we use 2 methods based on the optical (BPT and WHAN), the third method based on mid--IR (WISE) is different and confirms and reinforces our results.

The fact that AGN dominate in void interiors and walls could serve 
to obtain well traced voids in large cosmological volumes. 
This could be important not only for void statistics and galaxy 
evolution, which is itself interesting, but also for cosmological applications such as the AP test. Taking into account the upcoming new generation of galaxy surveys, these results can be relevant for several large scale studies.

\section*{Acknowledgments}
This work has been partially supported by Consejo de Investigaciones 
Cient\'{\i}ficas y T\'ecnicas de la Rep\'ublica Argentina (CONICET), the
Secretar\'{\i}a de Ciencia y T\'ecnica de la Universidad Nacional de C\'ordoba (SeCyT) and Secretar\'ia de Ciencia y T\'ecnica de la Universidad Nacional de San Juan.             
Funding for the SDSS and SDSS-II has been provided by the Alfred P. Sloan Foundation, the Participating Institutions, the National Science Foundation, the U.S. Department of Energy, the National Aeronautics and Space Administration, the Japanese Monbukagakusho, the Max Planck Society, and the Higher Education Funding Council for England. The SDSS Web Site is http://www.sdss.org/. The SDSS is managed by the Astrophysical Research Consortium for the Participating Institutions. The Participating Institutions are the American Museum of Natural History, Astrophysical Institute Potsdam, University of Basel, University of Cambridge, Case Western Reserve University, University of Chicago, Drexel University, Fermilab, the Institute for Advanced Study, the Japan Participation Group, Johns Hopkins University, the Joint Institute for Nuclear Astrophysics, the Kavli Institute for Particle Astrophysics and Cosmology, the Korean Scientist Group, the Chinese Academy of Sciences (LAMOST), Los Alamos National Laboratory, the Max-Planck-Institute for Astronomy (MPIA), the Max-Planck-Institute for Astrophysics (MPA), New Mexico State University, Ohio State University, University of Pittsburgh, University of Portsmouth, Princeton University, the United States Naval Observatory, and the University of Washington.
This research has used the NASA's Astrophysics Data System. 
Plots were performed using R software.

\section*{DATA AVAILABILITY}
The data underlying this article will be shared on reasonable request to the corresponding author.
\bibliographystyle{mnras}
\bibliography{agnsvoids.bib}
\end{document}